\documentclass{article}
\usepackage{graphicx} 
\usepackage{todonotes}
\usepackage{xspace}
\usepackage{enumitem}
\usepackage{xcolor}
\usepackage{tabularx}     
\usepackage{booktabs}     
\usepackage{subcaption}
\usepackage{hyperref}
\usepackage{tikz}
\usepackage{float}
\usepackage{booktabs}
\usepackage{geometry}
\usepackage{amsmath}
\usepackage{caption}
\usepackage{tabularx}
\usepackage{array}   
\usepackage{booktabs}
\usepackage{array}
\usepackage[table]{xcolor}
\usepackage{comment}
\usepackage{fancyhdr}

\definecolor{lightgray}{gray}{0.95}

\newcommand{\NetMob}{\texttt{NetMob26}\xspace}

\title{The NetMob26 Dataset: A High-Resolution Multi-Source \\View of Public Bus Mobility in Niterói}

\author{
  Felipe Domingos\textsuperscript{1}, Humberto T. Marques-Neto\textsuperscript{1}, Bruno Pereira\textsuperscript{2}, \\
  Clayson Celes\textsuperscript{3},Steffen Knoblauch \textsuperscript{4}, Vinícius F. S. Mota\textsuperscript{5} \thanks{These authors contributed equally to this work.}\\
   \vspace{0.2cm}
   \textsuperscript{1}PUC Minas, Brazil \quad
  \textsuperscript{2}UFBA, Brazil \quad
  \textsuperscript{3}ITA, Brazil
  \quad \\
  \textsuperscript{4} HeiGIT, Germany 
  \quad
  \textsuperscript{5}UFES, Brazil 
}
\date{May 2026}

\begin{document}

\maketitle

\begin{abstract}
High-quality data on public transportation systems and passenger demand remain limited, despite increasing interest from researchers and urban stakeholders in understanding how humans move and how transit services can be improved. The \NetMob Data Challenge addresses this gap by releasing a unique dataset capturing the dynamics of public transit mobility in Niterói. Unlike traditional survey-based datasets, the \NetMob dataset relies on operational data to provide a view of mobility, based on real-world records of bus fleet movements and passenger interactions during the month of March 2026.

The \NetMob dataset integrates four complementary data sources: (i) a mobility dataset containing GPS telemetry from the public bus fleet, with vehicle positions recorded approximately every 15–30 seconds, (ii) a ticketing dataset comprising around 7.2 million passenger boarding transactions, including fare types, anonymized user identifiers, and contextual temporal attributes, (iii) an auxiliary dataset including bus route geometries, bus stops, and meteorological information, and (iv) an urban infrastructure and socio-demographic dataset, containing parking lots, school and hospital locations. Together, these datasets provide a multi-layered representation of both the supply and demand aspects of urban mobility.

Before release, the data underwent preprocessing, cleaning, and anonymization procedures to ensure privacy while preserving analytical value. In particular, all passenger identifiers were anonymized, and operational inconsistencies, such as telemetry gaps and out-of-service vehicle records, were addressed to improve data quality. Access to the \NetMob dataset is granted to participants upon acceptance of the challenge’s Terms and Conditions and the signing of a Non-Disclosure Agreement (NDA).

This paper presents the data collection pipeline, dataset structure, preprocessing methodology, and an empirical characterization of mobility patterns observed in the system. The \NetMob dataset provides a valuable resource for research on public transportation efficiency, demand forecasting, accessibility analysis in Niterói, enabling studies on ridership patterns, service reliability, demand–supply interactions, and the impact of external factors such as weather on the local transit system.
\end{abstract}

\section{Introduction}
The rapid proliferation of mobile devices and location-aware technologies has led to the continuous generation of large-scale, high-resolution data on human activities and movement patterns. These data offer substantial opportunities for the analysis of urban systems~\cite{du2025review,khodabandelou2019estimation}, including public transport planning~\cite{seppecher2021zonal}, spatial inequality analyses~\cite{moro2021mobility,ucar2021socioeconomic}, and studying societal impacts of technological infrastructures~\cite{denadai2016death}.
 
In public transportation research, mobility datasets—such as GPS traces collected from bus fleets and passenger ticketing transactions—play a fundamental role in analyzing and improving system efficiency and operations. In addition, standardized formats such as GTFS and GTFS-Realtime (GTFS-RT) enable transportation networks to become ``computable'' by formally representing routes, stops, schedules, and real-time vehicle information. This standardization is essential for enabling reproducible analyses and facilitating comparisons across different cities and transit systems~\cite{aemmer2022measurement,lopes2024towards}. These data support a wide range of disciplines, including geography, sociology, computer science, public health, and economics, enabling detailed analyses of phenomena such as commuting dynamics~\cite{ma2017commuting}, service accessibility~\cite{bittencourt2023evaluating}, exposure to environmental conditions~\cite{nyhan2019airpollution}, and even the spread of infectious diseases~\cite{yabe2022mobile, knoblauch2025dengue}.

Over the last decade, researchers have increasingly relied on mobile network data as a scalable alternative to traditional surveys, enabling analyses at spatial and temporal resolutions that were previously difficult to achieve. Despite its clear benefits, access to openly accessible mobility data remains limited. Concerns about privacy, proprietary constraints, and the strategic importance of such data often restrict sharing by mobile operators and digital platforms, ultimately limiting innovation and undermining the reproducibility of scientific studies.

To alleviate these limitations, open data challenges have played a pivotal role in advancing methods and enabling large-scale empirical studies. Pioneering efforts such as the \textit{Data for Development} (D4D) challenges by Orange in collaboration with NetMob (2012–2014)~\cite{blondel2013d4d,montjoye2014d4dsenegal}, the ITU Ebola challenge (2015)~\cite{itu_bigdata}, the Telecom Italia Big Data Challenge (2014–2015)~\cite{barlacchi2015multisource}, the Data for Refugees (D4R) challenge in Turkey (2018)~\cite{salah2018d4r}, and the Future Cities Challenge supported by Foursquare at NetMob 2019~\cite{foursquare2018future}, have collectively shaped a vibrant research landscape around anonymized mobility data.

More recently, the \texttt{NetMob} conference series has revived this tradition through curated data releases that set new standards of accessibility and scientific utility. \texttt{NetMob23}~\cite{martinez2023netmob23} provided an extensive dataset of mobile application usage across $20$ metropolitan areas in France, mapped at a $100\times100$ m spatial resolution and covering $68$ popular mobile services. \texttt{NetMob24}~\cite{worldbank2024netmob} shifted its focus to countries in the Global South, offering aggregated mobility statistics from Mexico, Colombia, Indonesia, and India, aligned with Sustainable Development Goals and local policy needs.

Building on this trajectory, the \texttt{NetMob25}~\cite{netmob25} Data Challenge introduced a new dataset on human mobility in the Île-de-France region, derived from high-resolution GNSS tracking data. The data combined detailed GPS traces, validated trip records, and rich individual-level information, enabling analyses of travel behavior, multimodality, and daily mobility patterns. This edition reinforced the importance of integrating multiple data sources to better capture the complexity of urban mobility systems and further expanded research opportunities.

For the 2026 edition, the \NetMob Data Challenge releases a dataset on public transportation mobility in Niterói, capturing both supply and demand dynamics of the public bus network. The released \NetMob dataset integrates three complementary data perspectives of mobility: (i) GPS trajectories from the public bus fleet, providing detailed information on vehicle movements along routes recorded in 15-30 second intervals; (ii) structured data on bus lines; and (iii) anonymized smart card transaction records, representing passenger boardings across the system. Unlike survey-based datasets, \NetMob reflects real-world mobility behavior by capturing millions of events generated during daily transit operations. 

The \NetMob dataset supports a wide range of potential applications, including:
\begin{itemize}
    \item Analysis of passenger demand patterns across space and time;
    \item Modeling and prediction of bus travel times, delays, and reliability;
    \item Optimization of routes, schedules, and fleet allocation;
    \item Detection of anomalies and disruptions in transit operations;
    \item Study of accessibility and spatial coverage of public transportation;
    \item Investigation of equity and service provision in urban mobility systems.
\end{itemize}

By combining operational data with detailed network information, \NetMob provides researchers with a multi-layered dataset that enables data-driven insights into public transportation systems. It preserves user privacy while offering high analytical value for understanding and improving urban mobility.

The \NetMob dataset will be made available to participants upon acceptance of the challenge’s Terms and Conditions and the signing of a Non-Disclosure Agreement (NDA). This paper presents the data collection methodology, the structure of the released databases, and the anonymization and data-cleaning procedures. In addition, it provides an initial characterization of public transportation mobility patterns in Niterói, aiming to support its use across multiple applications.

\section{Niterói City and its Bus Transportation System}

\begin{figure}[htbp] 
\centering 
\includegraphics[width=1\textwidth]{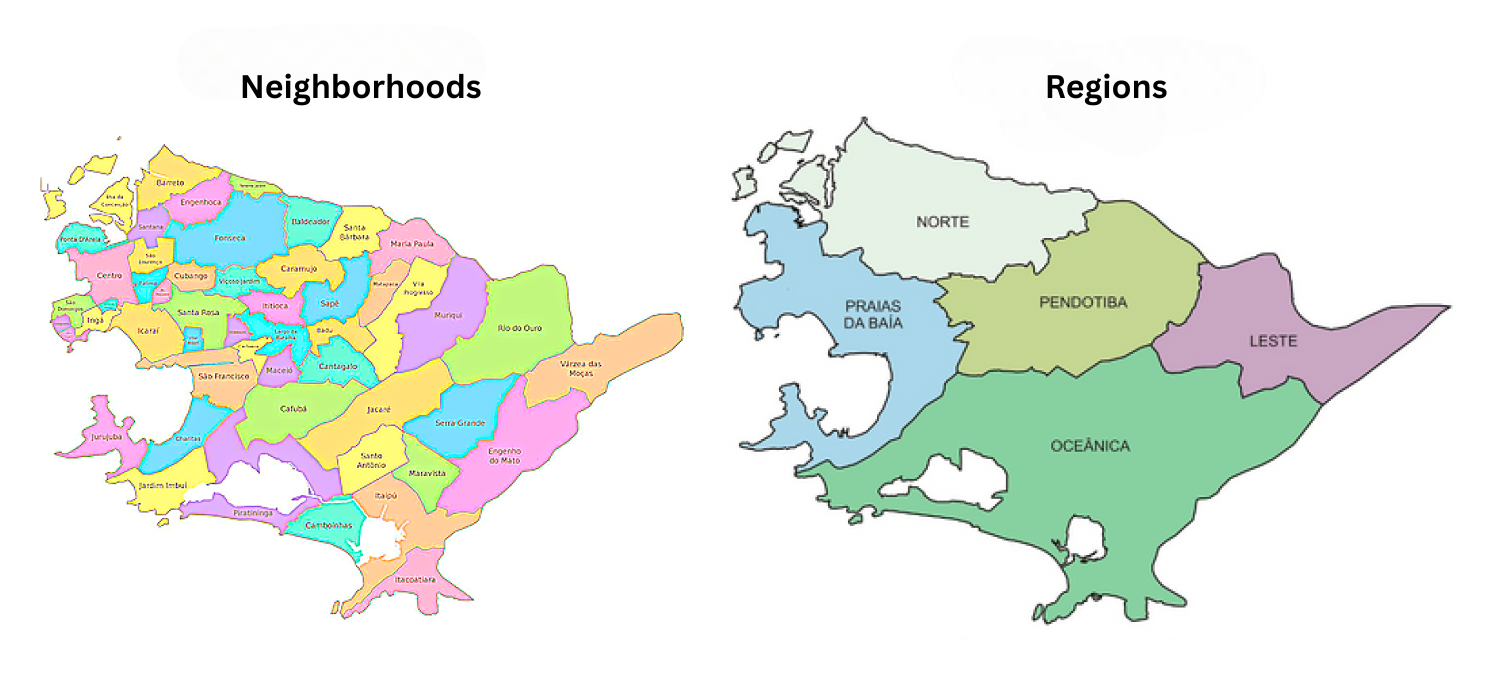} 
\caption{Administrative regions and neighborhoods of Niterói, Rio de Janeiro~\cite{edra2022catalogo}.} 
\label{fig:mapa_niteroi} 
\end{figure}

Niterói is a Brazilian city located in the state of Rio de Janeiro (RJ), Brazil. It has a total area of $133,757$ km² (2025) and a population of $481,749$~\cite{ibge_censo_2022}. The population density was around $3,601.67$ inhabitants/km² in 2022. Of its inhabitants, $231,427$ hold formal jobs (2023) and receive an average of $2.8$ times the minimum Brazilian salary (2023)~\cite{ibge_censo_2022}. The GDP per capita is R\$$158,314.82$ (2023), the human development index is $0.837$ (2022), and the life expectancy is $76.2$ years~ \cite{ibge_censo_2022}. The city of Niterói is an important metropolitan area of the municipality of Rio de Janeiro. These two urban areas are connected by the large Rio-Niterói bridge. The spatial distribution of the municipality is organized into $5$ (five) distinct administrative regions encompassing a total of $52$ (fifty-two) neighborhoods, as depicted in the thematic mapping of the city (see Figure \ref{fig:mapa_niteroi}). This territorial subdivision is a cornerstone for the political-administrative planning of Niterói, which stands as a substantial economic and industrial hub within the metropolitan region of Rio de Janeiro \cite{edra2022catalogo}.

\subsection{Geoinformation Management System – SIGeo}
\label{sec:SIGeo}

The municipality of Niterói, Brazil, operates the Geoinformation Management System (\textit{Sistema de Gestão da Geoinformação} – SIGeo)~\cite{sigeo2024}, a georeferenced urban data management platform developed and maintained by its Municipal Secretariat of Science, Technology, and Innovation since 2018. SIGeo integrates the acquisition, storage, analysis, and dissemination of spatial data to support evidence-based decision-making across domains such as urban planning, public health, environmental management, transportation, and public safety. The platform is structured around three core components: a centralized geographic database; a public web-based Geoportal enabling spatial queries, geoanalysis, and multi-format map exports; and an Open Data Portal that exposes dataset downloads, statistical indicators, and REST APIs to facilitate interoperability with third-party systems. 

The SIGeo Open Data Portal~\footnote{https://www.sigeo.niteroi.rj.gov.br/pages/dados-abertos, in Portuguese.} organizes its publicly available georeferenced datasets into 18 thematic categories, covering a broad spectrum of urban and territorial dimensions relevant to smart city research and municipal management. These categories span domains typically associated with urban computing applications -- including land use, infrastructure, mobility, environmental conditions, and socioeconomic indicators -- reflecting the municipality's commitment to structured, machine-readable spatial data dissemination.

To complement the SIGeo urban layers with fine-grained population structure, we incorporate census data from IBGE's 2022 demographic census, aggregated at the neighborhood level for the municipality of Niterói-RJ. The \textit{Instituto Brasileiro de Geografia e Estatística} (IBGE) is Brazil's official federal agency for statistical and geographic information, responsible for conducting national censuses and disseminating socioeconomic and demographic data across all administrative levels. 

\subsection{Bus Transportation System}
\label{sec:BUN}

The ``Bilhete Único Niterói'' (BUN) is a public policy instrument designed to enhance urban mobility and promote socio-economic integration within the municipality of Niterói. By implementing a fare integration system, the municipal government aims to reduce the financial burden on commuters who rely on multiple modes of transport to complete their daily journeys~\cite{riobilheteunico2024}. The technical operation of the BUN is governed by specific temporal and sequential constraints. According to the official guidelines provided by the management consortium~\cite{riobilheteunico2024}, the benefit allows for the integration of two distinct transport legs within a maximum time window of one hour ($60$ minutes). This integration applies to municipal buses in Niterói, enabling users to pay a unified, subsidized fare instead of the combined cost of two individual tickets.

To qualify for the tariff benefit, the following criteria must be met:
\begin{enumerate}
    \item \textbf{Card Registration}: the physical electronic card must be duly registered and linked to the user's Individual Taxpayer Registry (CPF).
    \item \textbf{Sequential Use}: the integration is valid for a maximum of two trips, provided they are not in the same direction on the same bus line (preventing round-trip misuse on a single fare).
    \item \textbf{Intermodal Scope}: while primarily focused on municipal bus lines, the system acts as a specialized branch of the broader state-level integration, though its specific Niterói-centric rules apply to journeys strictly within the municipal boundaries~\cite{riobilheteunico2024}.
\end{enumerate}

The implementation of this system was designed to facilitate access to the labor market and essential services for low-income populations residing in peripheral areas. By capping transport expenditure per journey, the BUN functions as a direct subsidy for citizens' mobility, effectively increasing the geographic reach of the urban population~\cite{riobilheteunico2024}.

\section{Data Collection}
Understanding mobility through public transportation is key to effective urban planning policy. Accurate and timely transit data is challenging to obtain and often expensive. The \NetMob Data Challenge provides a dataset of Niterói's bus system, leveraging operational data to support informed decision-making. 

\textbf{Scope and objectives.} The mobility dataset, covering \textbf{March 2026}, aims to provide a snapshot of urban mobility in Niterói. It integrates real-time bus GPS telemetry with passenger boarding transactions, capturing millions of trips and interactions. About \textbf{7.2 million boarding records} and over \textbf{370 thousand distinct users} (anonymized) illustrate urban transit dynamics across varied payment methods.

\textbf{Data collection and integration.} Mobility data was collected through the Mobnit API and follows the GTFS (General Transit Feed Specification) standard, with additional extensions to support real-time vehicle positioning attributes. Each bus reports its position approximately every \textbf{15–30 seconds}, enabling the reconstruction of detailed vehicle trajectories across the network. These data are complemented by structured information on routes, trip identifiers, and operational attributes.

The ticketing dataset records boarding events with timestamps, anonymized user IDs, fare types, vehicle and route information, and context such as time of day and day type. Combining these datasets allows joint analysis of \textbf{vehicle movements and passenger demand}, enabling insights into system performance and usage.

\textbf{Data processing and quality assurance.} Both datasets underwent preprocessing and cleaning. Bus telemetry was organized into daily CSV files, with each record representing a vehicle observation. Ticketing data was arranged into daily files. User information was anonymized to preserve privacy.

Operational data limitations persist. GPS telemetry may be incomplete due to connectivity issues, and vehicles without route information may indicate out-of-service movement. In ticketing, a user ID of zero indicates cash payments or unregistered users, while a zero fare value denotes free rides or transfers.

\textbf{Spatial and temporal coverage.} The dataset covers the urban area of Niterói within a bounding box of approximately [-22.95, -22.75] latitude and [-43.20, -42.95] longitude, including key transit hubs such as Barcas Terminal, Icaraí, Centro, and Piratininga. All timestamps are provided in BRT (UTC-3). The data captures a full month of operations, encompassing weekdays and weekends, enabling the analysis of temporal patterns such as peak demand, daily rhythms, and variability in transit usage.

\textbf{IBGE Census Data 2022.} The IBGE demographic dataset provides neighborhood-level population statistics for the municipality of Niterói, based on the 2022 Brazilian Demographic Census. The dataset aggregates demographic information for all 52 neighborhoods of the city, including total population counts and age-group distributions segmented by sex. Organized through standardized census variables, the data enables the characterization of the socioeconomic and demographic structure of urban regions and supports analyses related to mobility demand, accessibility, and transportation equity.

\textbf{Complementary social-demographic data.} To enrich the mobility analysis with a broader urban context, additional geospatial datasets were sourced from the SIGeo Open Data Portal (see Section~\ref{sec:SIGeo}), at the beginning of \textbf{May 2026}. From the available thematic categories, we selected a subset of layers directly related to urban mobility and its socioeconomic determinants, namely \textbf{education} facilities, \textbf{health} infrastructure, and \textbf{mobility}-related assets, such as parking lots and bicycle infrastructure. 

%Additionally, population and demographic data from the Brazilian national census, produced by IBGE, were integrated at the neighborhood granularity. This layer provides resident counts and age-sex distributions for all $52$ neighborhoods of Niterói-RJ, supporting equity analyses that correlate transit accessibility with the spatial distribution of age-sensitive groups, such as children, the working-age population, and the elderly.
\section{Dataset Overview and Anonymization}
The dataset released as part of the \NetMob Data Challenge provides a comprehensive, multi-source representation of public transportation mobility in Niteroi. Unlike survey-based datasets centered on individuals, \NetMob captures system-level mobility dynamics by integrating operational and transactional data collected in March 2026. The dataset comprises complementary components that, while not directly linked at the individual-trajectory level, can be combined to analyze the interaction between transport supply and passenger demand.

%%----------------------------------------------------------------------------------------------------------------
\subsection{Database Structure}

The dataset is organized into five groups of datasets, listed below. The main datasets are the mobility and ticketing, containing bus trajectory and ticket payment. The remaining datasets aim to enrich the information, allowing a more in-depth analysis of mobility in large cities.  
\begin{enumerate}

    \item \textbf{Mobility dataset (bus telemetry):} Contains high-frequency GPS observations of public buses, with positions recorded approximately every \textbf{15–30 seconds}. Each record includes vehicle identifiers, timestamps, route information (GTFS-based), geographic coordinates, and operational attributes such as direction and destination. These data enable the reconstruction of vehicle trajectories and the analysis of service performance.
    
    \item \textbf{Ticketing dataset (passenger transactions):} Contains approximately \textbf{7.2 million boarding records}, each representing a passenger interaction with the system. The dataset includes anonymized user identifiers, fare types (e.g., free, card, cash), vehicle and route information, integration flags (transfers), and temporal context (day type and time period). This dataset reflects \textbf{observed demand patterns} across the transit network.
    
    \item \textbf{Auxiliary dataset:} Includes static and contextual information such as route geometries (\texttt{line\_routes.json}), bus stop locations (\texttt{stops.json}), integration terminals, and meteorological data (temperature, precipitation, wind). These data support spatial and environmental analyses of mobility patterns.

    The file \texttt{meteorological\_data.csv} contains hourly weather observations for Niterói, including temperature, wind, and precipitation, covering the period from \textbf{March 1 to March 31, 2026}. The data were obtained from INMET and enable the analysis of environmental impacts on public transportation demand and operations.
    
    \item \textbf{IBGE Census Data 2022:} The IBGE Census 2022 dataset for Niterói provides spatial boundaries and socioeconomic indicators derived from the official Brazilian demographic census. The data are organized into thematic groups covering demographic composition, household infrastructure, literacy, urban environment characteristics, and household head profiles at different spatial resolutions, including census tracts, neighborhoods, and favelas/urban communities.

    %VM AQUI ESTA DANDO DETALHES QUE OS OUTROS NAO DAO
    %The file \texttt{population\_indicators\_2022.csv}, available in the \texttt{census\_tracts\_indicators/} directory, contains demographic indicators aggregated at the Census Tract level for the municipality of Niterói-RJ (IBGE municipality code 3303302). The dataset includes population counts disaggregated by age groups and sex, following the official IBGE Census 2022 methodology. Variable definitions and semantic descriptions are documented in the accompanying dictionary file \texttt{dict\_census\_tracts\_2022.csv}, which maps the original Portuguese indicator codes to their corresponding English descriptions while preserving the official IBGE nomenclature for consistency.

    The dataset structure follows the standardized organization adopted throughout the repository, where each record represents a Census Tract (\textit{Setor Censitário}) and can be spatially linked to the geographic boundaries available in the file \texttt{niteroi\_census\_borders.gpkg}. This integration enables spatial analysis combining demographic indicators with urban infrastructure, mobility, environmental, and socioeconomic data across the city of Niterói.

    %%The file \texttt{IBGE\_Aggregated\_Demo graphic\_Data\_by\_Neighborhood\_in\_Niteroi.csv} provides neighborhood-level demographic summaries for all 52 districts of Niterói-RJ (IBGE municipality code 3303302), with a total enumerated population of around 481,000 residents. The schema comprises 38 fields: two administrative identifiers (\texttt{Neighborhood Code}, a 10-digit national census tract code; \texttt{Neighborhood Name}, the neighborhood name in Portuguese) and 36 integer-valued demographic counts (variables \texttt{V01006}--\texttt{V01041}). Variable semantics are fully documented in the accompanying \texttt{IBGE\_Reduced\_Data\_Dictionary.csv}, which maps each code to its Portuguese label, thematic group (\textit{Demografia}), and entity type (\textit{Pessoas}); the broader \texttt{IBGE\_Complete\_ Data\_Dictionary.csv} provides the full 1,411-variable reference covering housing, mortality, literacy, and racial composition for researchers requiring additional census dimensions. Neighborhood-level linkage with SIGeo layers is achieved through the \texttt{Neighborhood Name} identifier, consistent with the \texttt{Neighborhood Text} attribute used across the urban infrastructure files.

    \item \textbf{ Points of Interest and Urban Infrastructure Datasets:} This component consists of multiple georeferenced CSV files sourced from the SIGeo Open Data Portal, categorized into \\ Education (e.g., \texttt{Education\_Universities.csv}; \texttt{Education\_Public\_Preschools.csv}), Health (e.g., \texttt{Healthy\_Basic\_Health\_Units.csv}), and Mobility (e.g., \texttt{Mobility\_Bicycle\_Stations. csv}; \texttt{Mobility\_Rotating\_Parking\_Spaces.csv}). The datasets follow a standardized schema, typically including spatial coordinates (X, Y projected in UTM Zone 23S), unique primary keys (\texttt{OBJECTID}, \texttt{globalid}, or \texttt{fid}), and descriptive attributes. 

\end{enumerate}

These components are \textbf{not keyed to a shared individual identifier} but can be integrated through \textbf{spatiotemporal alignment} (e.g., by associating boarding events with nearby stops and vehicle trajectories). This design reflects real-world data collection constraints while still enabling rich cross-dataset analyses.

The structural cohesion of the \NetMob database is established through the systematic integration of mobility trajectories with the authoritative urban layers described in the Urban Infrastructure and Socio-demographic Datasets.  These complementary social-demographic files provide a socio-spatial context required to correlate passenger demand with the availability of public services across Niterói-RJ. By utilizing shared attributes such as neighborhood identifiers and precise geographic coordinates, these datasets can be spatially aligned with the bus telemetry and ticketing records. This integrated database structure facilitates advanced urban computing research, enabling the empirical characterization of accessibility patterns and the evaluation of transit equity within the municipal transport network.

%%----------------------------------------------------------------------------------------------------------------
\subsection{Anonymization and Cleaning Procedures}

The anonymization process was designed to preserve the analytical value of the data while preventing the identification of individuals or sensitive patterns. The main steps of this pipeline are summarized as follows:

\begin{itemize}
    \item \textbf{User anonymization:} Passenger identifiers were irreversibly anonymized into the variable \texttt{anon\_user\_id}. Transactions with \texttt{anon\_user\_id = 0} correspond to cash payments or unregistered users. No personally identifiable information (e.g., names, card IDs) is present in the dataset.
    
    \item \textbf{Event-level data representation:} The dataset does not include reconstructed individual trajectories. Each record corresponds to a single boarding event, preventing continuous tracking of passengers across the network.
    
    \item \textbf{Operational data cleaning:} GPS telemetry data were filtered to remove incomplete or inconsistent records. Vehicles with missing \texttt{routeId} or \texttt{linha} are typically associated with out-of-service movements or depot transfers.
    
    \item \textbf{Handling missing and noisy data:} Occasional gaps in GPS traces may occur due to connectivity issues. These gaps were preserved but can be identified through temporal discontinuities. Spatial attributes (latitude, longitude, heading) are stored with high precision and can be optionally smoothed or aggregated for specific analyses.
    
    \item \textbf{Data consistency and standardization:} All datasets were organized into daily CSV files and validated to ensure temporal and spatial consistency. All timestamps are provided in BRT (UTC-3), and spatial data is constrained to the urban boundaries of Niteroi.
\end{itemize}

\subsection{Analytical Considerations}

Mobility and ticket datasets were preprocessed, cleaned, and organized into daily CSV files to enhance usability. Some considerations when using the data include:

\begin{itemize}
    \item \textbf{Telemetry gaps:} Possible due to connectivity issues;
    \item \textbf{Out-of-service vehicles:} Identified by missing route information;
    \item \textbf{Cash transactions:} Represented by anonymized user IDs equal to zero;
    \item \textbf{Zero fare values:} May indicate free rides or integrated transfers.
\end{itemize}

All timestamps are provided in \textbf{BRT (UTC-3)}, and the spatial coverage includes the main urban area of Niterói, encompassing key mobility hubs such as Barcas Terminal, Centro, Icaraí, and Piratininga.

\section{Database Structure and Variables}

This section describes the structure and content of the core datasets released in the \NetMob Data Challenge. For each dataset, we present the main variables and their semantics, and explain how they can be combined to support comprehensive analyses of public transportation systems. A more detailed description of the datasets and their preprocessing steps is available in the official documentation.

%%------------------------------------------------------------------------------------------------------------
\subsection{Mobility Dataset (Bus Telemetry)}

The mobility dataset contains high-frequency GPS observations of \textbf{public buses} operating in Niterói. Each row corresponds to a single vehicle observation, typically recorded every 15–30 seconds, enabling the reconstruction of detailed trajectories and operational patterns.

Table~\ref{tab:vehicle_schema} presents the key variables of the dataset and their descriptions.

\begin{table}[htb]
\centering
\begin{tabular}{lp{10cm}}
\toprule
\textbf{Variable} & \textbf{Description} \\
\midrule
\texttt{id}        & Unique vehicle identifier \\
\texttt{timestamp} & Observation time (BRT, UTC-3) \\
\texttt{tripId}    & GTFS trip identifier encoding schedule and direction \\
\texttt{routeId}   & GTFS route identifier \\
\texttt{lat}       & Geographic latitude (WGS84) \\
\texttt{lng}       & Geographic longitude (WGS84) \\
\texttt{angle}     & Vehicle heading in degrees (0\textdegree = North, 90\textdegree = East) \\
\texttt{linha}     & Public-facing route code (e.g., 31, 49.1) \\
\texttt{nomeLinha} & Full route name (origin--destination) \\
\texttt{headsign}  & Destination displayed on the vehicle \\
\texttt{sentido}   & Trip direction (e.g., outbound/inbound) \\
\bottomrule
\end{tabular}
\caption{Vehicle Telemetry Data Schema}
\label{tab:vehicle_schema}
\end{table}

This dataset supports analyses of vehicle movement, service reliability, route performance, and spatiotemporal dynamics of the transit system.

%%-------------------------------------------------------------------------------------------------------------
\subsection{Ticketing Dataset (Passenger Transactions)}
The ticketing dataset contains individual boarding events from the bus system, representing the demand side of urban mobility. Each record corresponds to a single passenger transaction collected during \textbf{March 2026}, totaling approximately \textbf{7.2 million events}.

Table~\ref{tab:transaction_schema} presents the key variables of the dataset and their descriptions. This dataset enables the analysis of ridership patterns, demand distribution, fare structures, and transfer behaviors across the transit network.

\begin{table}[ht]
\centering

\begin{tabular}{lp{10cm}}
\toprule
\textbf{Variable} & \textbf{Description} \\
\midrule
\texttt{transaction\_date} & Boarding timestamp (BRT, UTC-3) \\
\texttt{anon\_user\_id}    & Anonymized passenger identifier \\
\texttt{fare\_type}        & Payment modality (free, card, or cash) \\
\texttt{vehicle\_number}   & Bus identifier \\
\texttt{company\_number}   & Operator identifier \\
\texttt{route\_detail\_id} & Route identifier \\
\texttt{route\_name}       & Full route name \\
\texttt{integration\_flag} & Indicates transfers between trips \\
\texttt{card\_type}        & Passenger category (e.g., student, senior) \\
\texttt{debited\_amount}   & Brazilian Reais (BRL) \\
\texttt{day\_type}         & Day classification (weekday/weekend) \\
\texttt{day\_period}       & Time-of-day classification \\
\bottomrule
\end{tabular}
\caption{Passenger Transaction Data Schema.}
\label{tab:transaction_schema}
\end{table}

%%-------------------------------------------------------------------------------------------------------------
\begin{comment}
    
\subsection{Integrated Mobility Perspective}

Unlike traditional survey-based datasets, \NetMob does not rely on self-reported trips or individual tracking devices. Instead, it provides a system-level view of mobility, combining:
\begin{itemize}
    \item \textbf{Supply-side data:} Bus trajectories and operational characteristics;
    \item \textbf{Demand-side data:} Passenger boarding transactions.
\end{itemize}
\end{comment}

%%-------------------------------------------------------------------------------------------------------------
\subsection{Auxiliar Dataset}

To support the analyses, the dataset also includes a set of auxiliary files containing detailed information about the public transport infrastructure. Among these resources, the geospatial files \texttt{line\_routes.json} and \texttt{stops.json} describe the complete network topology by providing the geometric representations of bus routes and the official street-level locations of bus stops, respectively. The integration files \texttt{stops\_integration\_city.json} and \texttt{stops\_integration\_metropolitan.json} are selected bus stops containing city integration terminals and the locations of major metropolitan interchange hubs. Finally, \texttt{meteorological\_data.csv} provides hourly weather conditions, recorded throughout March 2026,  from the official Brazilian Meteorology Institute (INMET) database. Each record includes measurements related to temperature, atmospheric pressure, wind conditions, solar radiation, and precipitation.

\begin{table}[htb]
    \centering
    \caption{Auxiliary Data Files}
    \label{tab:auxiliary_data}
    \rowcolors{2}{gray!10}{white}
    \begin{tabular}{>{\ttfamily}p{6cm} p{8cm}}
        \toprule
        \textbf{File} & \textbf{Description} \\
        \midrule
        \texttt{line\_routes.json}                     & Complete geometric shapes of all bus routes. \\
        \texttt{stops.json}                            & Official bus stops with street-level location. \\
        \texttt{stops\_integration\_city.json}         & Real-time snapshots near city integration terminals. \\
        \texttt{stops\_integration\_metropolitan.json} & Locations of major metropolitan interchange hubs. \\
        \texttt{meteorological\_data.csv}              & Hourly weather data (Temp, Rain, Wind) during March 2026 from the official Brazilian Meteorology Institute (INMET)\\
        \bottomrule
    \end{tabular}
\end{table}

\begin{comment}
Table~\ref{tab:weather_schema} presents the key variables of the dataset and their descriptions.

\begin{table}[ht]
\centering
\begin{tabular}{lp{10cm}}
\toprule
\textbf{Variable} & \textbf{Description} \\
\midrule
\texttt{timestamp\_utc}   & Observation timestamp in UTC \\
\texttt{temp\_ins}        & Instantaneous temperature (°C) \\
\texttt{temp\_max}        & Maximum temperature during the interval (°C) \\
\texttt{temp\_min}        & Minimum temperature during the interval (°C) \\
\texttt{pressure\_ins}    & Instantaneous atmospheric pressure (hPa) \\
\texttt{pressure\_max}    & Maximum atmospheric pressure (hPa) \\
\texttt{pressure\_min}    & Minimum atmospheric pressure (hPa) \\
\texttt{wind\_speed}      & Wind speed (m/s) \\
\texttt{wind\_dir}        & Wind direction (degrees) \\
\texttt{wind\_gust}       & Wind gust intensity (m/s) \\
\texttt{radiation}        & Solar radiation (kJ/m$^2$) \\
\texttt{rain}             & Precipitation (mm) \\
\bottomrule
\end{tabular}
\caption{Meteorological Data Schema.}
\label{tab:weather_schema}
\end{table}
\end{comment}
This auxiliary dataset enables analysis of bus stop predictions, traffic in major hubs,  and environmental impacts on urban mobility, including the effects of rainfall, temperature variations, and wind conditions on service reliability and operational efficiency.

%%------------------------------------------------------------------------------------------------------------

%%------------------------------------------------------------------------------------------------------------
\subsection{Point of Interest and Urban Infrastructure Dataset} 
\label{sec:data_urbaninfrasctructure_dataset}

To contextualize the mobility dynamics within the socioeconomic landscape of Niterói-RJ, a suite of georeferenced datasets was integrated from the authoritative \textit{Sistema de Gestão da Geoinformação} (SIGeo) Open Data Portal~\cite{sigeo2024}. 

\begin{table}[htb]
    \centering
    \caption{Social Data Files and Descriptions}
    \label{tab:social_data}
     \rowcolors{2}{gray!10}{white}
    \begin{tabular}{>{\ttfamily}p{6cm} p{8cm}}
        \toprule
        \normalfont\textbf{File} & \textbf{Description} \\
        \midrule
        Health\_Regional.csv      & Boundaries of regional health administrative areas. \\
        Health\_Healthcare\_Service\_Area.csv         & Service-area delimitations for public\_healthcare\_coverage. \\
        Health\_Hospitals.csv                       & Locations and attributes of hospitals. \\
        Health\_Polyclinics.csv                     & Locations and attributes of polyclinics. \\
        Health\_Basic Health Units.csv              & Locations and attributes of basic health units (UBS). \\
        Health\_Pharmacies.csv                      & Locations and attributes of pharmacies. \\
        Education\_Universities.csv                 & Locations and attributes of universities. \\
        Education\_State\_Middle\_Schools.csv         & Locations and attributes of state-run middle schools. \\
        Education\_Municipal\_Primary\_Schools.csv    & Locations and attributes of municipal primary schools. \\
        Education\_Public\_Preschools.csv            & Locations and attributes of public preschools. \\
        Mobility\_Private\_Parking\_Lots.csv          & Locations and attributes of private parking lots. \\
        Mobility\_Rotating\_Parking\_Spaces.csv       & On-street rotating (paid) parking spaces. \\
        Mobility\_Bicycle Stations.csv              & Locations of bicycle-sharing stations. \\
        Garbage Collection Schedule.csv             & Schedule and routes for municipal garbage collection. \\
        Restaurants\_2019.csv                       & Restaurant inventory and attributes for 2019. \\
        \hiderowcolors
        \bottomrule
    \end{tabular}
\end{table}

All of these datasets are available in Portuguese (but easily translatable) at \url{www.sigeo.niteroi.rj.gov.br/pages/dados-abertos}, and they are organized into thematic groups that reflect the city's essential services and infrastructure. We organized a subset of SIGeo’s datasets, especially with relevant data of \textbf{Education}, \textbf{Health}, and \textbf{Mobility},  supplemented by contextual layers such as \texttt{Restaurants\_2019.csv} and \texttt{Garbage\_Collection\_Schedule.csv}. This multi-layered approach enables researchers to correlate passenger boarding patterns with urban attractors and public service availability across the municipality's 52 neighborhoods.

The datasets follow a standardized CSV structure derived from ArcGIS Feature Layers, with a relatively consistent attribute naming convention. Each record is uniquely identified through primary keys such as \texttt{OBJECTID}, \texttt{globalid}, or \texttt{fid}, which facilitate relational joins across different urban layers. Spatially, the data is projected in \texttt{UTM Zone 23S} using the \texttt{SIRGAS 2000} datum, ensuring precise alignment with the bus telemetry trajectories. The released CSV files include explicit \texttt{X} and \texttt{Y} projected coordinates to enhance interoperability for non-GIS specialized environments.

A full description of the dataset can be found in Appendix \ref{appendix:poi}.

%Data quality and consistency were rigorously validated against the official SIGeo metadata. Consistency checks inspired by technical audits revealed that while the majority of attributes align with municipal standards, certain discrepancies exist between the physical CSV files and the ArcGIS templates. For instance, in the \texttt{Mobility\_Private\_Parking\_Lots.csv} dataset, the field \texttt{tx\_site} is used in place of the standard \texttt{tx\_website}, and certain layers – such as \texttt{Healthy\_Basic\_Health\_Units.csv} – exhibit occasional \texttt{globalid} duplication that requires deduplication during the ingestion phase. Furthermore, missing values in metadata fields such as \texttt{tx\_email} or \texttt{tx\_telefone} are preserved as null strings to preserve the integrity of the original administrative records.

%The integration of these thematic layers with the core mobility dataset is primarily achieved through administrative neighborhood identifiers or through direct spatial joins based on the provided coordinates. This structural cohesion enables evaluation of transit equity by mapping the proximity of boardings to specific facilities. By adhering to open data principles of accessibility and integration, these datasets provide the necessary socio-spatial backbone for advanced urban computing tasks, including accessibility modeling and evaluating public service areas.

%%------------------------------------------------------------------------------------------------------------
\subsection{Social Demographic Census IBGE Dataset} 
\label{sec:data_socialdemographic_dataset}

This dataset contains spatial boundaries and census indicators for the city of Niterói, derived from the 2022 Census conducted by the \textit{Instituto Brasileiro de Geografia e Estatística} (IBGE). Folder and file names were standardized in English to facilitate international use and interoperability.

\noindent\textbf{Directory Structure:}

\begin{table}[h]
\centering
\caption{Main directory structure of the IBGE Census dataset.}
\label{tab:ibge_structure}
\begin{tabular}{>{\ttfamily}p{0.38\linewidth} p{0.55\linewidth}}
\toprule
\normalfont\textbf{Filename} & \textbf{Description} \\
\midrule
niteroi\_census\_borders.gpkg & GeoPackage containing spatial boundaries (census tracts and neighborhoods). \\
favelas\_indicators/ & Statistics related to favelas and urban communities in Niterói. \\
census\_tracts\_indicators/ & Statistical indicators aggregated at the Census Tract level. \\
\bottomrule
\end{tabular}
\end{table}

\begin{table}[h]
\centering
\caption{Files available in the \texttt{census\_tracts\_indicators} directory.}
\label{tab:census_tracts_indicators}
\rowcolors{2}{gray!10}{white}
\begin{tabular}{>{\ttfamily}p{0.55\linewidth} p{0.45\linewidth}}
\toprule
\normalfont\textbf{Filename} & \textbf{Description} \\
\midrule
dict\_census\_tracts\_2022.csv & Metadata dictionary defining variable codes and labels for census tract indicators. \\
density\_indicators\_2022.csv & Population density indicators (e.g., residents per hectare and residents per household). \\
household\_indicators\_2022.csv & Absolute counts of household infrastructure characteristics by census tract. \\
population\_indicators\_2022.csv & Demographic profile aggregated by census tract. \\
literacy\_indicators\_2022.csv & Literacy and illiteracy counts for population groups within each tract. \\
household\_heads\_indicators\_2022.csv & Demographic characteristics of household heads by census tract. \\
block\_face\_environment\_indicators.csv & Absolute counts of urban infrastructure features observed on street/block faces. \\
block\_face\_environment\_indicators\_pct.csv & Percentage coverage of urban infrastructure features (e.g., sidewalks, ramps, lighting). \\
\bottomrule
\end{tabular}
\end{table}

\begin{table}[htb]
\centering
\caption{Files available in the \texttt{favelas\_indicators} directory.}
\label{tab:favelas_indicators}
\rowcolors{2}{gray!10}{white}
\begin{tabular}{>{\ttfamily}p{0.50\linewidth} p{0.50\linewidth}}
\toprule
%\rowcolor{gray!25}
\normalfont\textbf{Filename} & \textbf{Description} \\
\midrule
dict\_favelas\_census.csv & Metadata dictionary defining codes and labels for absolute count variables. \\
dict\_favelas\_census\_pct.csv & Metadata dictionary defining codes and labels for percentage-based variables. \\
household\_indicators\_2022.csv & Absolute counts of household infrastructure indicators (e.g., water supply, sanitation, waste disposal). \\
household\_indicators\_pct\_2022.csv & Percentage distribution of household infrastructure characteristics. \\
population\_indicators\_2022.csv & Demographic indicators including population counts by age group and sex. \\
population\_indicators\_pct\_2022.csv & Percentage distribution of demographic cohorts by age and sex. \\
literacy\_indicators\_2022.csv & Absolute counts of literate and illiterate individuals. \\
literacy\_indicators\_pct\_2022.csv & Literacy rate indicators across mapped communities. \\
household\_heads\_indicators\_2022.csv & Absolute counts of household heads categorized by sex, race, and age. \\
household\_heads\_indicators\_pct\_2022.csv & Percentage breakdown of household head demographics. \\
household\_heads\_disaggregated\_pct\_2022.csv & Fine-grained demographic cross-tabulations of household heads. \\
neighborhood\_environment\_indicators.csv & Absolute counts of urban surroundings features aggregated by neighborhood. \\
neighborhood\_environment\_indicators\_pct.csv & Percentage coverage of public infrastructure indicators by neighborhood. \\
\bottomrule
\end{tabular}
\end{table}

\newpage
\noindent\textbf{Common Header Translations (IBGE Terminology)}

When translating the metadata dictionaries (\texttt{dict\_*.csv}), the following terminology mappings are recommended:

\begin{description}
    \item[Setor Censitário] Census Tract
    \item[Domicílios] Households
    \item[Entorno] Urban Surroundings / Built Environment
    \item[Alfabetização] Literacy
    \item[Responsável pelo domicílio] Head of Household
    \item[Favelas e Comunidades Urbanas (FCU)] Favelas and Urban Communities (formerly Subnormal Agglomerations)
    \item[Face] Block Face / Street Segment
\end{description}

\begin{quote}
\textbf{Attention:} The indicator names contained in the dictionaries remain in Portuguese to preserve consistency with the original IBGE datasets. However, English descriptions are provided to facilitate interpretation by international users.
\end{quote}

\newpage 
\section{Exploratory Dataset Characterization}

This section presents an initial exploratory characterization of the datasets released in the \NetMob Data Challenge, providing an overview of their structure, scale, and key statistical properties. The analysis focuses on three complementary components of the urban mobility ecosystem in \textit{Niterói, RJ, Brazil}: bus mobility (supply), passenger transactions (demand), and meteorological conditions (context).

%%---------------------------------------------------------------------------------------------------------
\subsection{Bus Mobility Characterization}

The bus mobility dataset captures the operational dynamics of the public transportation system through GPS telemetry. Each vehicle reports its position, enabling the reconstruction of detailed trajectories across the urban network.

An initial characterization of this dataset focuses on understanding spatial coverage, temporal dynamics, and fleet behavior. Figures in this section illustrate the geographic distribution of vehicle positions, highlighting the main transit corridors and areas with higher bus density. These visualizations provide insight into the network's structure and the relative importance of different regions within the system.

\begin{figure}[htp]
    \centering
    \includegraphics[width=0.65\linewidth]{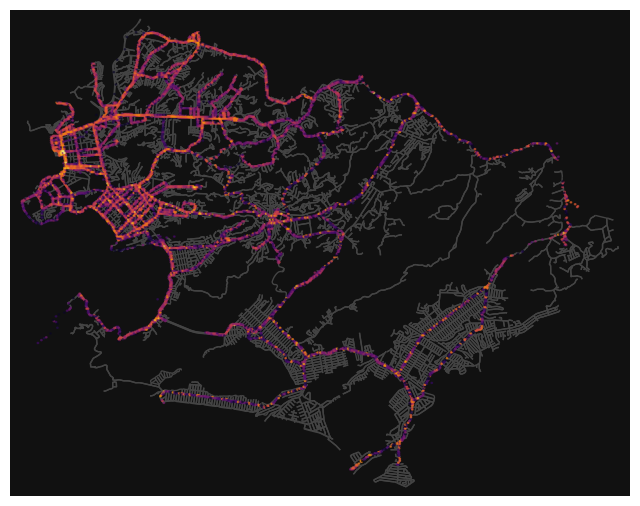}
    \caption{Spatial Characterization: Niterói Transit Density (March 11-14).}
    \label{fig:mapa01}
\end{figure}

Figure~\ref{fig:mapa01} presents the public transportation traffic patterns in Niterói between March 11 and 14, highlighting the most active mobility flows and the city’s main transit corridors.

\begin{table}[ht]
\centering
\caption{Trip Duration Statistics (Minutes).}
\label{tab:trip_duration_stats}
\begin{tabular}{lr}
\toprule
\textbf{Statistic} & \textbf{Value (min)} \\
\midrule
Mean    & $87.42$ \\
Std Dev & $42.12$ \\
Minimum & $2.02$ \\
25\%    & $59.82$ \\
50\% (Median) & $81.57$ \\
75\%    & $111.55$ \\
Maximum & $262.23$ \\
\bottomrule
\end{tabular}
\end{table}

\begin{figure}[htb]
    \centering
    \includegraphics[width=0.75\linewidth]{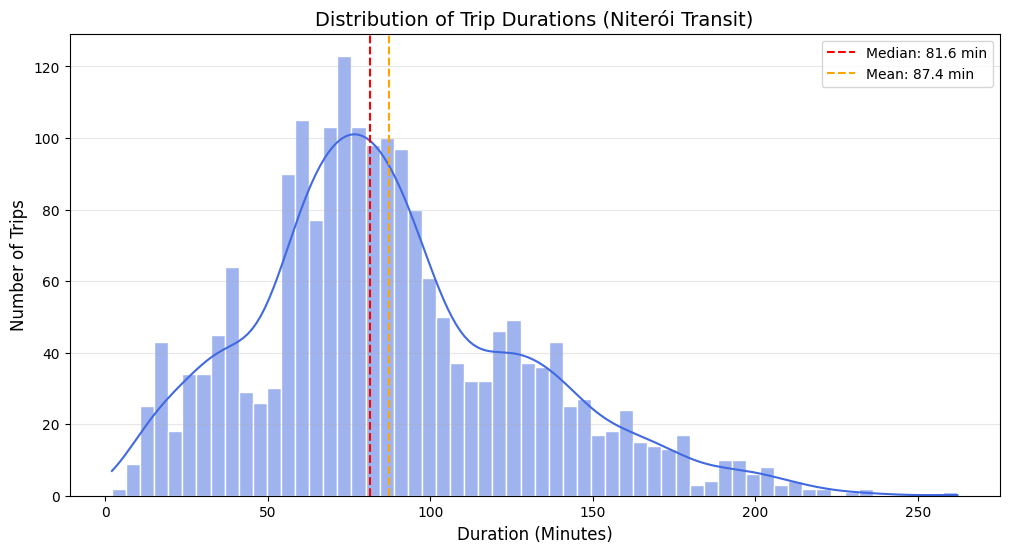}
    \caption{Distribution of Trip Durations.}
    \label{fig:duration}
\end{figure}

Table~\ref{tab:trip_duration_stats} and Figure~\ref{fig:duration} present the statistical summary and distribution of trip durations extracted from the public transportation dataset collected between March 11 and 14 in Niterói. The results indicate that most trips have moderate durations, with an average travel time of approximately $87.42$ minutes and a median of $81.57$ minutes, suggesting a slightly right-skewed distribution influenced by longer trips.

The standard deviation of $42.12$ minutes demonstrates a considerable variability in travel times, reflecting the heterogeneity of mobility patterns across different routes and urban regions during the analyzed period. The interquartile range, between $59.82$ minutes ($25$\%) and $111.55$ minutes ($75$\%), shows that the majority of trips are concentrated within this interval.

As illustrated in Figure~\ref{fig:duration}, the distribution presents a higher concentration of trips between approximately $40$ and $120$ minutes, while a smaller number of trips exhibit significantly longer durations, reaching a maximum of $262.23$ minutes. These longer trips may be associated with peripheral routes, peak-hour congestion, operational delays, or complex commuting patterns observed during the days analyzed.

\begin{figure}[htb]
    \centering
    \includegraphics[width=0.75\linewidth]{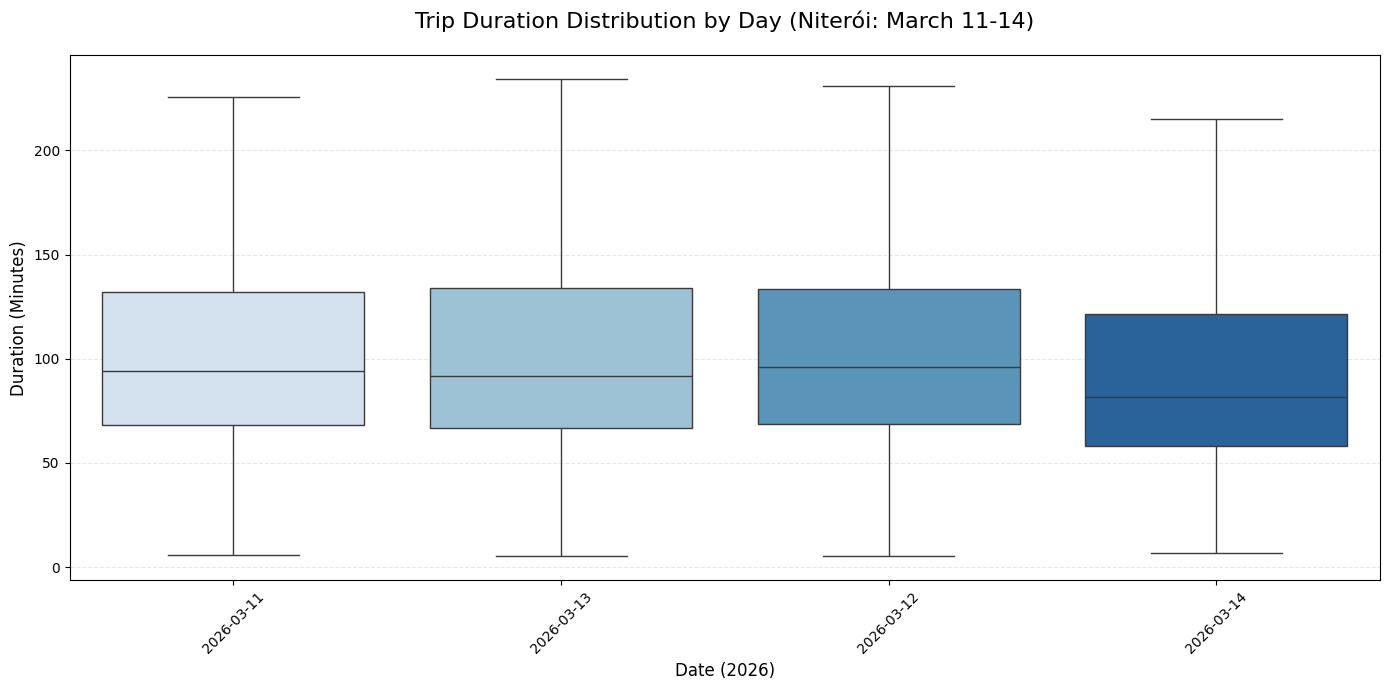}
    \caption{Trip Duration Distribution by Day (Niterói: March 11-14).}
    \label{fig:durationbyday}
\end{figure}

Figure~\ref{fig:durationbyday} shows the distribution of trip durations in Niterói between March 11 and 14, 2026. The dataset contains 1,606 trips on March 11, 1,595 on March 12, 1,339 on March 13, and 876 on March 14, indicating a reduction in mobility demand over the analyzed period.

The results reveal relatively stable trip duration patterns across all days, with median travel times between $80$ and $100$ minutes. March 12 and 13 present slightly higher durations, while March 14 shows lower median values and fewer trips, possibly reflecting different mobility behavior. Additionally, all days exhibit some long-duration trips above $200$ minutes, suggesting the presence of congestion, operational delays, or longer commuting routes.

\begin{figure}
    \centering
    \includegraphics[width=0.75\linewidth]{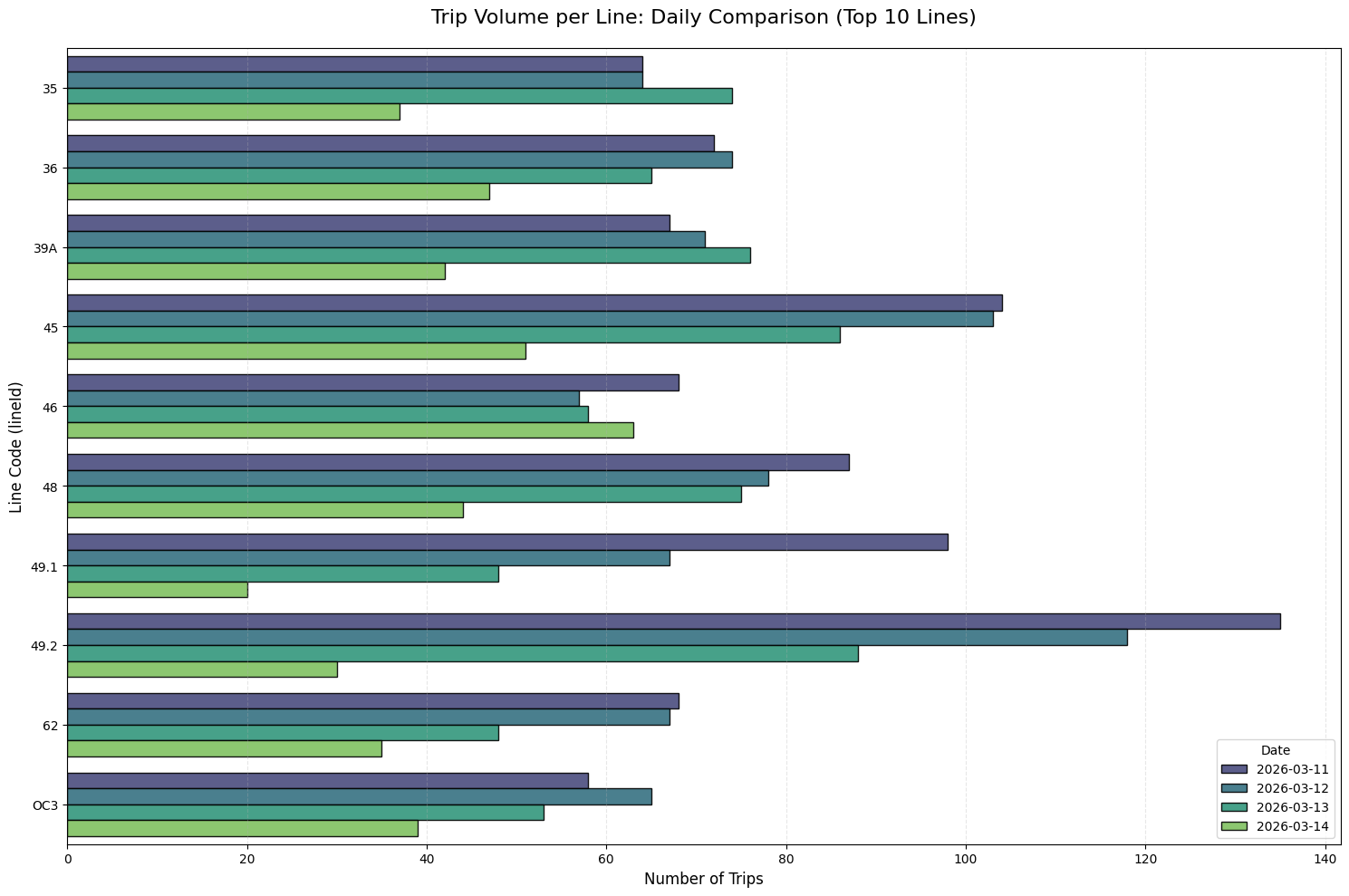}
    \caption{Trip Volume per Line: Daily Comparison (Top 10 Lines).}
    \label{fig:tripvolume}
\end{figure}

Figure~\ref{fig:tripvolume} shows the daily comparison of trip volumes for the top 10 public transportation lines in Niterói between March 11 and 14, 2026. The results indicate that line 49.2 consistently presented the highest number of trips, particularly on March 11 and 12, highlighting its importance as one of the city’s main transit corridors. Other lines, such as 45, 49.1, and 48, also exhibited high trip demand, while lines like 35 and 62 showed comparatively lower volumes throughout the analyzed period.

A noticeable reduction in trip counts can be observed on March 14, 2026, across almost all lines. Since this date corresponds to a Saturday, the decrease likely reflects typical weekend mobility behavior, characterized by lower commuting demand and changes in travel patterns. Despite this reduction, some routes, such as 46 and OC3, maintained relatively stable trip volumes, suggesting more regular transportation demand independent of weekday commuting flows. Overall, the results reveal heterogeneous mobility dynamics and show that a small subset of bus lines concentrates a significant portion of public transportation demand in Niterói.

%%---------------------------------------------------------------------------------------------------------
\subsection{Ticket Transactions Characterization}

The ticketing dataset reflects the demand side of the public transportation system, capturing individual boarding events across the network. With millions of transactions recorded over the observation period, the dataset enables a detailed analysis of passenger behavior and ridership patterns. The characterization begins with aggregate statistics on transaction volume, including distributions by payment type (e.g., free, card, cash) and user category. Visualizations highlight the relative contribution of each fare modality, providing insight into system usage and subsidy structures.

\begin{figure}[htb]
    \centering
    \includegraphics[width=0.9\linewidth]{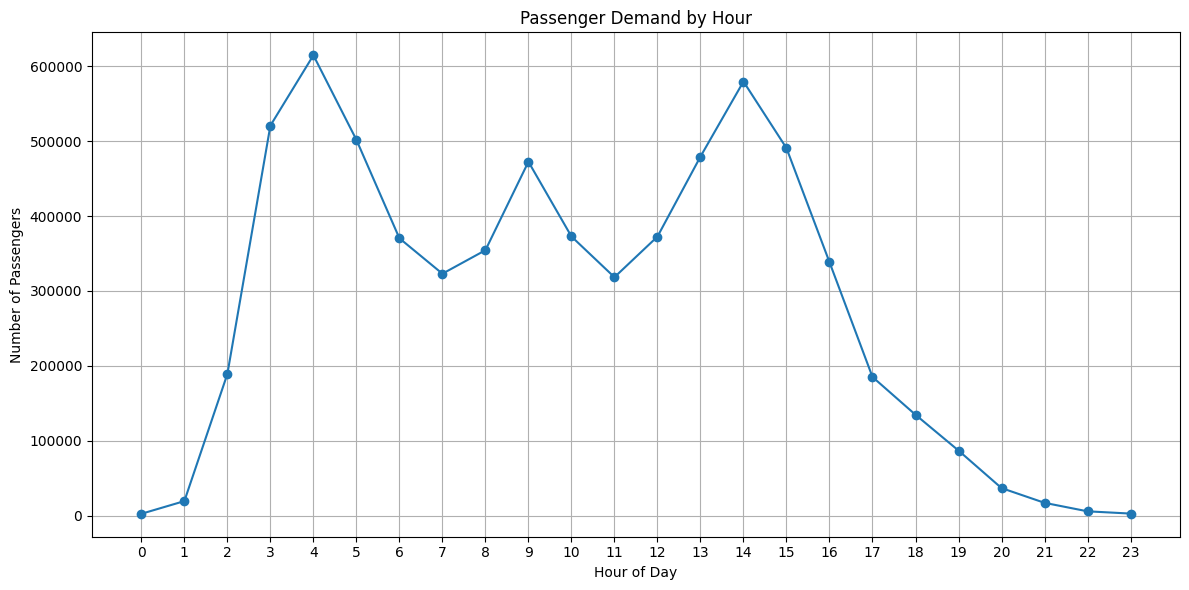}
    \caption{Hourly passenger demand distribution throughout the day.}
    \label{fig:hourly_passenger_demand}
\end{figure}

The graph presented in Figure~\ref{fig:hourly_passenger_demand} illustrates the hourly distribution of passenger demand throughout the day. Passenger volume is very low during late-night hours, reaching its minimum around midnight and 23:00. Demand starts to increase significantly after 02:00, with a sharp growth observed between 03:00 and 05:00, corresponding to the beginning of daily commuting activities.

The highest passenger demand occurs at 04:00, reaching approximately $614$ thousand passengers, followed by another significant peak at 14:00 with nearly $580$ thousand passengers. After the early morning peak, demand gradually decreases during the morning hours, stabilizes around midday, and rises again in the early afternoon. From 16:00 onward, passenger volume steadily declines until returning to very low levels during nighttime hours.

\begin{figure}[htb]
    \centering
    \includegraphics[width=0.95\linewidth]{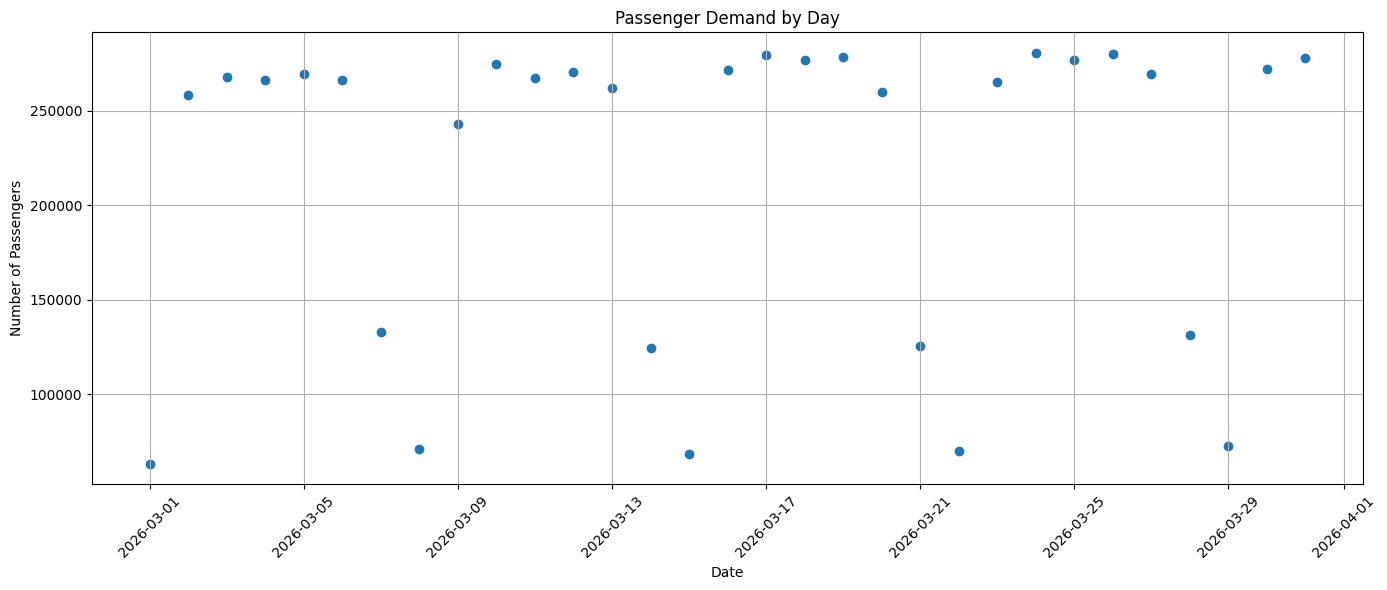}
    \caption{Daily passenger demand during March 2026.}
    \label{fig:daily_passenger_demand}
\end{figure}

The graph presented in Figure~\ref{fig:daily_passenger_demand} illustrates the daily variation in passenger demand throughout March 2026. A clear weekly mobility pattern can be observed, strongly associated with weekdays and weekends.

Passenger demand remains consistently high during weekdays (Monday to Friday), generally ranging between $240$ thousand and $280$ thousand passengers per day. The highest demand values are observed on weekdays such as March 17, March 24, March 26, and March 31, all exceeding $277$ thousand passengers. These elevated values are associated with regular commuting activities, including work, education, and commercial travel.

In contrast, Saturdays exhibit a noticeable reduction in passenger volume, with demand typically ranging between $124$ thousand and $133$ thousand passengers. Sundays present the lowest mobility levels of the week, with passenger counts between approximately $63$ thousand and $73$ thousand passengers. For example, March 1, March 8, March 15, March 22, and March 29 correspond to Sundays and show the lowest demand values in the dataset.

\begin{figure}[htb]
    \centering
    \includegraphics[width=0.95\linewidth]{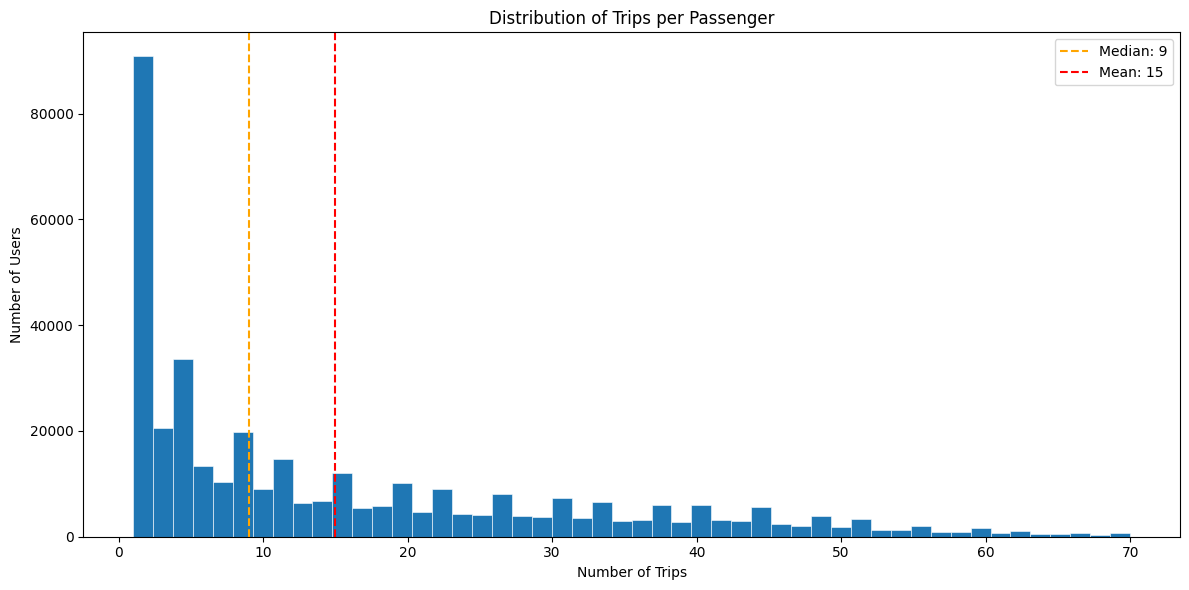}
    \caption{Distribution of trips per passenger during the observation period. The histogram highlights the asymmetry of passenger travel behavior, including the mean and median number of trips per user.}
    \label{fig:trips_per_passenger}
\end{figure}

Figure~\ref{fig:trips_per_passenger} presents the distribution of trip counts per passenger considering all identified users in the dataset, excluding anonymous cash transactions. The distribution exhibits a strongly right-skewed behavior, indicating that most passengers perform a relatively small number of trips, while a smaller group of users performs a substantially larger number of trips. The highest concentration of users is observed between 1 and 10 trips during the analyzed period, revealing that occasional or low-frequency users represent a significant portion of the passenger population.

The median number of trips per passenger is approximately $9$ trips, represented by the orange dashed line, while the mean reaches approximately $15$ trips, represented by the red dashed line. The mean being substantially higher than the median indicates the presence of heavy users who perform many trips and consequently pull the average upward.

Additionally, the long tail extending beyond $50$ and reaching nearly $70$ trips demonstrates the existence of highly recurrent passengers, likely associated with daily commuters who depend heavily on public transportation for work, education, or routine activities. Overall, the figure highlights the heterogeneity of passenger mobility behavior and reinforces the existence of distinct usage profiles within the transportation system.

%%----------------------------------------------------------------------------------------------------------
\subsection{Meteorological Analysis}

The meteorological dataset provides information on weather conditions during the data collection period, including variables such as temperature, precipitation, and wind intensity. Although external to the transportation system, these factors can influence mobility behavior and system performance. The characterization of this dataset focuses on identifying temporal patterns and variability in weather conditions throughout the observation period. Time-series plots illustrate daily and hourly fluctuations, highlighting periods of rainfall or extreme temperatures.

\begin{figure}[htb]
    \centering
    \begin{subfigure}[b]{0.45\linewidth}
        \centering
        \includegraphics[width=\linewidth]{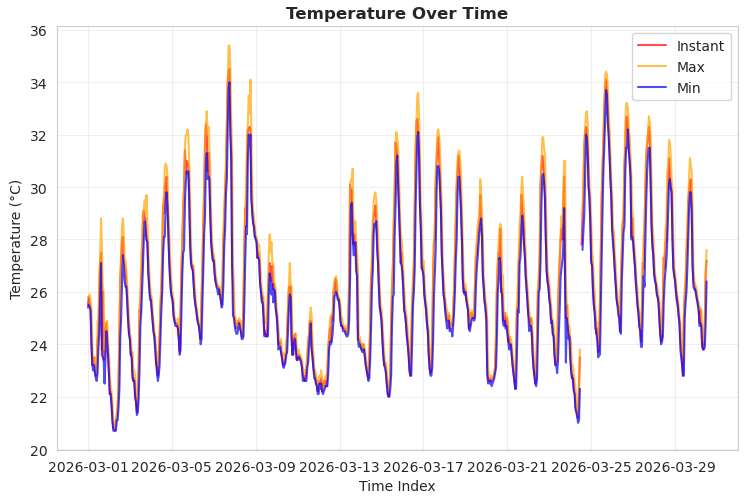}
        \caption{Temperature Over Time.}
        \label{fig:temp01}
    \end{subfigure}
    \hfill
    \begin{subfigure}[b]{0.45\linewidth}
        \centering
        \includegraphics[width=\linewidth]{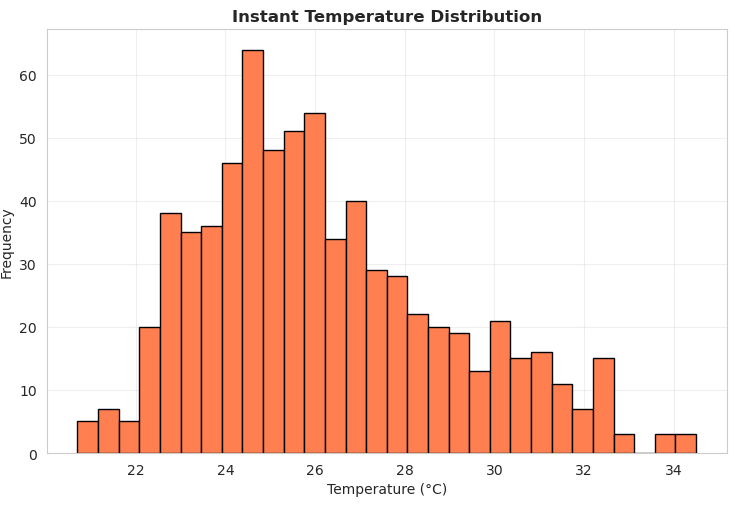}
        \caption{Instant Temperature Distribution.}
        \label{fig:temp02}
    \end{subfigure}
    \\
    \begin{subfigure}[b]{0.45\linewidth}
        \centering
        \includegraphics[width=\linewidth]{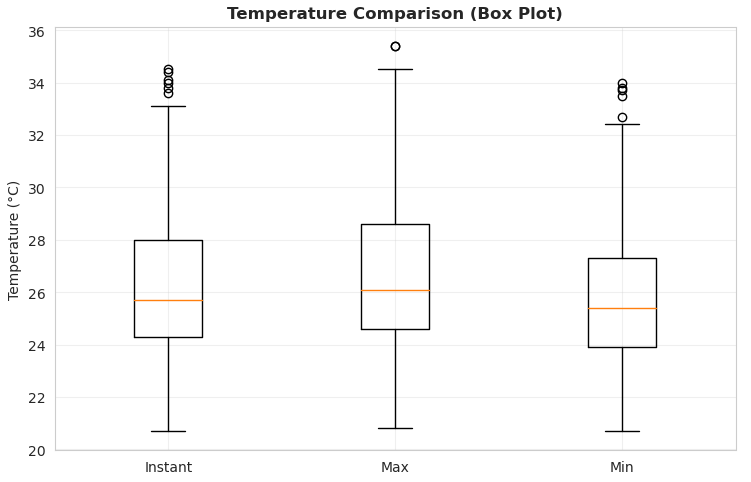}
        \caption{Temperature Comparison (Box Plot).}
        \label{fig:temp03}
    \end{subfigure}
    \caption{Temperature Analysis over March 2026.}
    \label{fig:temperature}
\end{figure}

The Figure~\ref{fig:temperature} presents a comprehensive analysis of temperature behavior in Niterói during March 2026, including temporal evolution, statistical distribution, and comparative variability between instant, maximum, and minimum temperatures. Figure~\ref{fig:temp01} illustrates the temperature variation over time throughout the month. The results reveal clear cyclical daily patterns, with recurrent peaks and valleys corresponding to daytime heating and nighttime cooling processes. Maximum temperatures frequently exceeded 30$^\circ$C, reaching peaks close to 35$^\circ$C, while minimum temperatures generally remained between 21$^\circ$C and 25$^\circ$C. Despite short-term fluctuations, the overall temperature behavior remained relatively stable during the analyzed period, reflecting typical tropical climatic conditions.

Figure~\ref{fig:temp02} presents the distribution of instant temperatures. The histogram indicates that most temperature observations are concentrated between 23$^\circ$C and 28$^\circ$C, with the highest frequency occurring around 25$^\circ$C to 26$^\circ$C. The distribution exhibits a slight right skew, indicating the occurrence of warmer periods with temperatures above 30$^\circ$C, although such events are less frequent.

Finally, Figure~\ref{fig:temp03} compares the statistical distributions of instant, maximum, and minimum temperatures using boxplots. The median temperatures are relatively similar across the three variables, ranging approximately from 25$^\circ$C to 26$^\circ$C. However, maximum temperatures present greater variability and higher upper extremes, reaching values above 35$^\circ$C. In contrast, minimum temperatures exhibit lower dispersion, suggesting more stable nighttime thermal conditions. The presence of outliers in all three distributions highlights sporadic extreme temperature events during the month.

\begin{figure}[htb]
    \centering
    \begin{subfigure}[b]{0.75\linewidth}
        \centering
        \includegraphics[width=\linewidth]{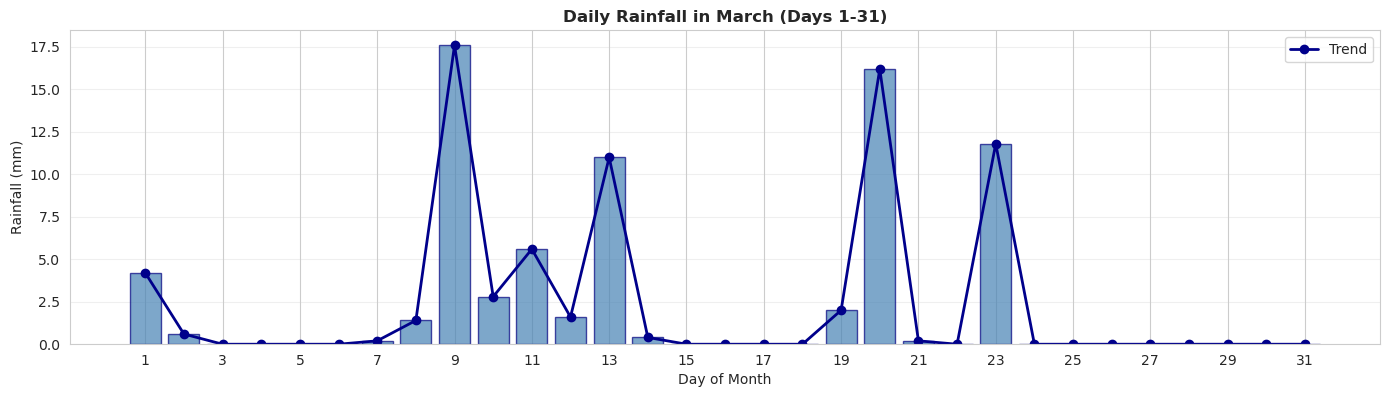}
        \caption{Daily Rainfall in March (Days 1-31).}
        \label{fig:rain01}
    \end{subfigure}
    \\
    \begin{subfigure}[b]{0.75\linewidth}
        \centering
        \includegraphics[width=\linewidth]{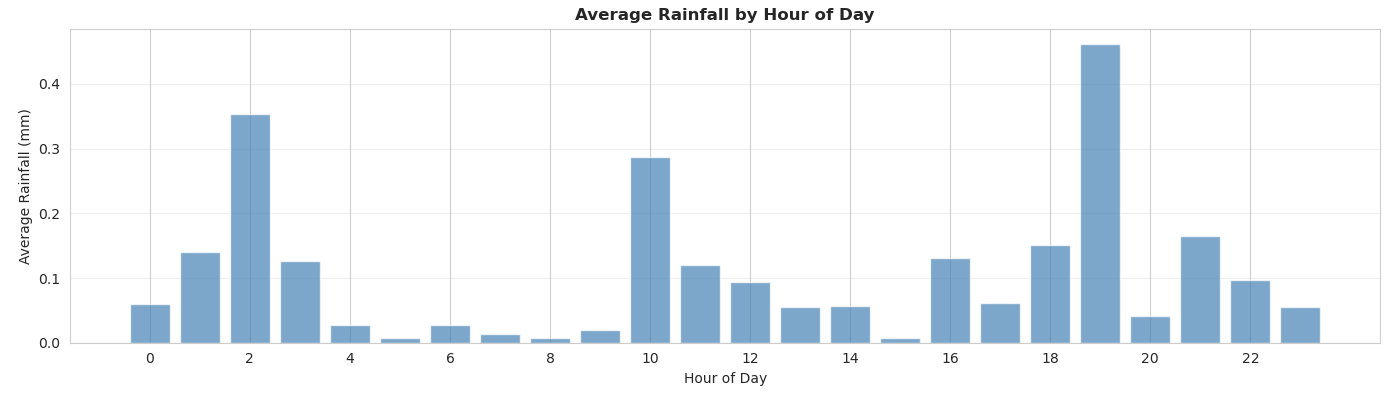}
        \caption{Average Rainfall by Hour of Day.}
        \label{fig:rain02}
    \end{subfigure}
    \caption{Daily Rainfall Summary - March 2026.}
    \label{fig:rain}
\end{figure}

Figure~\ref{fig:rain} presents the rainfall behavior in Niterói during March 2026, including both the daily rainfall accumulation and the hourly rainfall distribution. The results indicate that precipitation events were irregularly distributed throughout the month, with several days presenting no rainfall and a few days concentrating most of the accumulated precipitation.

As shown in Figure~\ref{fig:rain01}, rainfall peaks occurred mainly on March 9, March 20, and March 23, with daily accumulations of $17.6$ mm, $16.2$ mm, and $11.8$ mm, respectively. Additional significant rainfall events were observed on March 13 ($11.0$ mm) and March 11 ($5.6$ mm). Overall, the total rainfall accumulated during March reached $75.60$ mm, with an average daily rainfall of $2.44$ mm. Only 14 out of the 31 days recorded precipitation above zero, demonstrating the intermittent nature of rainfall during the analyzed period.

Figure~\ref{fig:rain02} illustrates the hourly rainfall distribution, revealing that precipitation intensity varied considerably throughout the day. The highest average rainfall was observed at 19:00, reaching approximately 0.46 mm, followed by elevated rainfall levels around 02:00 and 10:00. In contrast, early morning hours such as 05:00, 08:00, and 15:00 presented very low rainfall averages. These results suggest that rainfall events were more frequent during late afternoon, evening, and nighttime periods, which may influence urban mobility conditions, traffic congestion, and public transportation operation dynamics.

\section{Methods for Future Analyses}

This section outlines possible analytical directions enabled by the \NetMob dataset, along with practical guidelines for its appropriate use—particularly regarding data integration, temporal analysis, and spatial granularity.

%%--------------------------------------------------------------------------------------------------------------
\subsection{Analytical Potential}

The richness and scale of the dataset—combining \textbf{bus GPS telemetry, passenger transactions, and meteorological data}—enable a wide range of studies in transportation systems, urban mobility, and data-driven policy design in Niterói. Examples include:

\begin{itemize}
    \item \textbf{Operational analysis:} Estimating headways, service frequency, and fleet utilization across routes;
    \item \textbf{Demand modeling:} Inferring ridership patterns from boarding transactions and identifying peak demand periods;
    \item \textbf{Service efficiency:} Correlating operational metrics (e.g., delays, route deviations) with passenger demand and revenue;
    \item \textbf{Accessibility studies:} Evaluating spatial coverage of the bus network and identifying underserved regions;
    \item \textbf{Arrival time prediction:} Developing machine learning models using spatiotemporal GPS sequences;
    \item \textbf{Demand–supply alignment:} Mapping high-demand corridors by combining transaction data with vehicle trajectories;
    \item \textbf{Weather impact analysis:} Investigating how meteorological conditions (rain, temperature, wind) influence ridership and operations;
    \item \textbf{Equity and policy analysis:} Studying fare categories (e.g., student, senior, subsidized) and their distribution across the system.
\end{itemize}

\subsection{Further Documentation and Resources}

Additional technical details, schema descriptions, and usage examples are available in the dataset repository. In particular:
\begin{itemize}
    \item \texttt{README\_Mobility.md} — detailed description of GPS telemetry data;
    \item \texttt{README\_Ticket.md} — schema and categorical mappings for transactions;
    \item Reference notebooks — exploratory analyses for mobility, ticketing, and meteorological data.
\end{itemize}

Together, these materials provide a solid foundation for researchers to explore, analyze, and generate impactful insights on \textbf{public transportation systems and urban mobility dynamics}.
\section{Conclusion and Availability}

This report presented the \NetMob dataset, a comprehensive multi-source urban mobility dataset that captures the dynamics of public transportation in Niterói, Brazil, during March 2026. By integrating bus GPS telemetry, passenger ticketing transactions, meteorological observations, and urban infrastructure data, the dataset provides a rich and realistic representation of both the operational and demand dimensions of the city's transit system. In addition to detailing the database structure and variables, this document describes the collection process, preprocessing pipeline, anonymization strategies, and exploratory characterization analyses designed to support reproducible and privacy-preserving research.

Beyond its descriptive value, the \NetMob dataset establishes a foundation for a broad spectrum of studies in transportation systems, urban computing, machine learning, accessibility analysis, and public policy evaluation. Its multi-layered structure enables researchers to investigate complex interactions between mobility demand, service supply, environmental conditions, and socioeconomic factors. By combining operational and contextual perspectives, the dataset supports the development of predictive models, optimization strategies, and evidence-based approaches for improving urban mobility systems.

More broadly, the initiative reinforces the importance of open and well-documented mobility datasets for advancing scientific research and fostering innovation in smart cities. We expect the \NetMob Data Challenge to stimulate new methodologies, interdisciplinary collaborations, and practical solutions capable of supporting more efficient, equitable, and sustainable public transportation systems.

\paragraph{Dataset access.}
The dataset is accessible within the \NetMob Data Challenge framework, subject to acceptance of the terms and conditions available at the following link:\footnote{\url{https://netmob.org/www26/files/Challenge_NDA.pdf}}. Upon approval, participants can access the full dataset, reference notebooks, and associated documentation through the official repository:\footnote{\url{https://github.com/lprm-ufes/Netmob2026}}.

\bibliographystyle{plain}
\bibliography{sections/references}

\newpage
\appendix

\section{Point of Interest and Urban infrastructure dataset detail}
\label{appendix:poi}

The \NetMob \textbf{Neighborhood Bounding Box} dataset defines the official administrative boundaries of the 52 neighborhoods of Niterói, providing the geometric representation of each spatial unit through attributes such as area and perimeter. This dataset acts as the fundamental spatial reference layer for integrating the other \NetMob components — including health, education, mobility, and sociodemographic datasets — within a unified neighborhood-based geographic framework. Table~\ref{tab:neighborhood_bbox_schema} summarizes the main variables of the neighborhood bounding box dataset and their respective descriptions.

\begin{table}[h!]
\centering
\begin{tabular}{lp{10cm}}
\toprule
\textbf{Variable} & \textbf{Description} \\
\midrule
\texttt{Object ID 1}       & Sequential record index  \\
\texttt{Object ID}         & Official numeric identifier for the neighborhood  \\
\texttt{Name Text}         & Official name of the neighborhood \\
\texttt{Geometry Area}     & Surface area of the neighborhood polygon (in square meters) \\
\texttt{Geometry Length}   & Perimeter length of the neighborhood polygon (in meters) \\
\texttt{Legislation Text}  & Municipal legislation under which the neighborhood boundary is formally defined (Lei Urbanística N\textsuperscript{o} 3905/2024) \\
\bottomrule
\end{tabular}
\caption{Neighborhood Bounding Box Dataset Schema.}
\label{tab:neighborhood_bbox_schema}
\end{table}

The \NetMob \textbf{Education} dataset of Niterói comprises four educational datasets, each capturing a distinct layer of the public and private education network across the municipality's neighborhoods. 

\noindent \textbf{Public Preschools}: covers the early childhood education units serving children from 4 months to 5 years and 11 months, organized by age group into nursery (berçário, 0–3 years) and preschool (pré-escola, 4–5 years) stages; nationally, ~78\% of preschool enrollments are in public institutions and ~22\% in private ones. Table~\ref{tab:preschools_schema} presents the key variables of the public preschools dataset and their descriptions.

\begin{table}[htb]
\centering
\begin{tabular}{lp{10cm}}
\toprule
\textbf{Variable} & \textbf{Description} \\
\midrule
\texttt{X Coordinate}    & Projected X coordinate (easting) of the school's location \\
\texttt{Y Coordinate}    & Projected Y coordinate (northing) of the school's location \\
\texttt{Object ID}       & Unique numeric identifier for each record \\
\texttt{School Text}     & Official name of the preschool institution \\
\texttt{Type Text}       & Administrative type of the unit (e.g., UMEI, NAEI) \\
\texttt{Shift Text}      & Operating shift (e.g., Integral, Parcial) \\
\texttt{Modality Text}   & Educational modality offered \\
\texttt{Years 2022 Text} & Age range or grade levels served in 2022 \\
\texttt{Address Text}    & Full street address of the school \\
\texttt{Neighborhood Text} & Neighborhood where the school is located \\
\texttt{Region Text}     & Administrative region of the municipality \\
\texttt{Telephone Text}  & Contact telephone number(s) \\
\texttt{Email Text}      & Official contact email address \\
\texttt{Global ID}       & Globally unique identifier (GUID) for the record \\
\bottomrule
\end{tabular}
\caption{Public Preschools Dataset Schema.}
\label{tab:preschools_schema}
\end{table}

\vspace{+6pt}

\noindent \textbf{Municipal Primary Schools}: the municipally managed schools offering the full range of fundamental education (1st to 9th grade, ages 6–14), including regular primary cycles (EF), early childhood education (EI), and youth and adult education (EJA); in the initial years (1st–5th grade), the municipal network is dominant with ~70\% of enrollments, followed by private schools (~19\%) and state schools (~11\%).
Table~\ref{tab:primary_schools_schema} presents the key variables of the municipal primary schools dataset and their descriptions.

\begin{table}[htb]
\centering
\begin{tabular}{lp{10cm}}
\toprule
\textbf{Variable} & \textbf{Description} \\
\midrule
\texttt{X Coordinate}      & Projected X coordinate (easting) of the school's location \\
\texttt{Y Coordinate}      & Projected Y coordinate (northing) of the school's location \\
\texttt{Object ID}         & Unique numeric identifier for each record \\
\texttt{School Text}       & Official name of the municipal primary school \\
\texttt{Type Text}         & Type of education offered (e.g., Ensino Fundamental) \\
\texttt{Shift Text}        & Operating shift (e.g., Parcial) \\
\texttt{Modality Text}     & Educational modalities offered (e.g., EI, EF, EJA) \\
\texttt{Years 2022 Text}   & Grade levels or age ranges served in 2022 \\
\texttt{Address Text}      & Full street address of the school \\
\texttt{Neighborhood Text} & Neighborhood where the school is located \\
\texttt{Region Text}       & Administrative region of the municipality \\
\texttt{Telephone Text}    & Contact telephone number(s) \\
\texttt{Email Text}        & Official contact email address \\
\texttt{Global ID}         & Globally unique identifier (GUID) for the record \\
\bottomrule
\end{tabular}
\caption{Municipal Primary Schools Dataset Schema.}
\label{tab:primary_schools_schema}
\end{table}

\noindent \textbf{State Middle Schools}: catalogs the state-administered institutions covering the final years of lower secondary education (6th to 9th grade, ages 11–14); across Brazil, the state network accounts for ~40\% of enrollments in these grades, the municipal network for ~44\%, and private schools for ~16\%. 
Table~\ref{tab:middle_schools_schema} presents the key variables of the state middle schools dataset and their descriptions.

\begin{table}[htb]
\centering
\begin{tabular}{lp{10cm}}
\toprule
\textbf{Variable} & \textbf{Description} \\
\midrule
\texttt{X Coordinate}        & Projected X coordinate (easting) of the school's location \\
\texttt{Y Coordinate}        & Projected Y coordinate (northing) of the school's location \\
\texttt{Object ID 1}         & Secondary numeric identifier for each record \\
\texttt{Object ID}           & Primary unique numeric identifier for each record \\
\texttt{Name Text}           & Official name of the state middle school \\
\texttt{Address Text}        & Street name of the school's address \\
\texttt{Address Number Text} & Street number of the school's address \\
\texttt{Telephone Text}      & Contact telephone number \\
\texttt{Neighborhood Text}   & Neighborhood where the school is located \\
\bottomrule
\end{tabular}
\caption{State Middle Schools Dataset Schema.}
\label{tab:middle_schools_schema}
\end{table}

\noindent \textbf{Universities}: compiles the higher education institutions present in the municipality, serving students typically aged 18 and above across undergraduate and postgraduate programs; higher education falls under federal responsibility, encompassing both undergraduate degrees and technology programs lasting more than two years, with a share of enrollments in private institutions nationally.
Table~\ref{tab:universities_schema} presents the key variables of the universities dataset and their descriptions.

\begin{table}[htb]
\centering
\begin{tabular}{lp{10cm}}
\toprule
\textbf{Variable} & \textbf{Description} \\
\midrule
\texttt{X Coordinate}        & Projected X coordinate (easting) of the institution's location \\
\texttt{Y Coordinate}        & Projected Y coordinate (northing) of the institution's location \\
\texttt{Object ID}           & Unique numeric identifier for each record \\
\texttt{Name Text}           & Official name of the higher education institution \\
\texttt{Address Text}        & Street name of the institution's address \\
\texttt{Address Number Text} & Street number of the institution's address \\
\texttt{Neighborhood Text}   & Neighborhood where the institution is located \\
\texttt{Telephone Text}      & Contact telephone number \\
\texttt{Email Text}          & Official contact email address \\
\texttt{ZIP Code Text}       & Postal (ZIP) code of the institution \\
\texttt{Website Text}        & Official website URL \\
\bottomrule
\end{tabular}
\caption{Universities Dataset Schema.}
\label{tab:universities_schema}
\end{table}

The \NetMob \textbf{Health} dataset of Niterói comprises five health datasets, each capturing a distinct layer of the public and private healthcare infrastructure across the municipality's neighborhoods.

\noindent \textbf{Regional Health Boundaries}: defines the six administrative health regions into which the municipality of Niterói is divided — Leste Oceânica, Norte I, Norte II, Pendotiba, Praias da Baía I, and Praias da Baía II — providing the spatial boundaries used to organize and coordinate health service delivery across the territory. Table~\ref{tab:health_boundaries_schema} presents the key variables of the regional health boundaries dataset and their descriptions.

\begin{table}[htb]
\centering
\begin{tabular}{lp{10cm}}
\toprule
\textbf{Variable} & \textbf{Description} \\
\midrule
\texttt{Feature ID}    & Unique numeric identifier for each health region record \\
\texttt{Health Region} & Name of the administrative health region \\
\texttt{Global ID}     & Globally unique identifier (GUID) for the record \\
\texttt{Shape Length}  & Perimeter length of the region's boundary polygon (in projected units) \\
\texttt{Shape Area}    & Surface area of the region's boundary polygon (in projected units) \\
\bottomrule
\end{tabular}
\caption{Regional Health Boundaries Dataset Schema.}
\label{tab:health_boundaries_schema}
\end{table}

\noindent \textbf{Basic Health Units} (Unidades Básicas de Saúde — UBS): catalogs the primary care facilities distributed across Niterói's neighborhoods, representing the first point of contact between residents and the public health system; UBS units provide preventive care, health monitoring, and basic clinical services, and are organized under municipal administration linked to specific health regions. Table~\ref{tab:basic_health_units_schema} presents the key variables of the basic health units dataset and their descriptions.

\begin{table}[htb]
\centering
\begin{tabular}{lp{10cm}}
\toprule
\textbf{Variable} & \textbf{Description} \\
\midrule
\texttt{Feature ID}               & Unique numeric identifier for each record \\
\texttt{CNES Code}                & National Registry of Health Establishments (CNES) code \\
\texttt{Name}                     & Official name of the basic health unit \\
\texttt{Generic Name}             & Common or abbreviated name of the unit \\
\texttt{Unit Type}                & Type of health establishment (e.g., Unidade Básica de Saúde) \\
\texttt{Care Level}               & Level of healthcare provided (e.g., Primária) \\
\texttt{Scale}                    & Administrative scale of the unit (e.g., Municipal) \\
\texttt{Street Name}              & Street name of the unit's address \\
\texttt{Number}                   & Street number of the unit's address \\
\texttt{Address Complement}       & Additional address details (e.g., floor, suite) \\
\texttt{ZIP Code}                 & Postal (ZIP) code of the unit \\
\texttt{Neighborhood}             & Neighborhood where the unit is located \\
\texttt{Municipality}             & Municipality where the unit is located \\
\texttt{State}                    & State where the unit is located \\
\texttt{Country}                  & Country where the unit is located \\
\texttt{Telephone}                & Contact telephone number \\
\texttt{Primary Health Care Team} & Indicates whether a primary health care team is assigned \\
\texttt{Health Region 2}          & Administrative health region the unit belongs to \\
\texttt{Global ID}                & Globally unique identifier (GUID) for the record \\
\texttt{Nickname Text}            & Informal or alternative name associated with the unit \\
\bottomrule
\end{tabular}
\caption{Basic Health Units Dataset Schema.}
\label{tab:basic_health_units_schema}
\end{table}

\noindent \textbf{Polyclinics} (Policlínicas Regionais): covers the regional polyclinics that serve as secondary-level care hubs within each health region, offering specialized outpatient consultations, diagnostic services, and referral support to complement the primary care provided by UBS units; each polyclinic is linked to a specific health region and operates under municipal administration. Table~\ref{tab:polyclinics_schema} presents the key variables of the polyclinics dataset and their descriptions.

\begin{table}[htb]
\centering
\begin{tabular}{lp{10cm}}
\toprule
\textbf{Variable} & \textbf{Description} \\
\midrule
\texttt{Feature ID}          & Unique numeric identifier for each record \\
\texttt{Telephone}           & Contact telephone number \\
\texttt{Neighborhood}        & Neighborhood where the polyclinic is located \\
\texttt{Municipality}        & Municipality where the polyclinic is located \\
\texttt{Number}              & Street number of the polyclinic's address \\
\texttt{Street Name}         & Street name of the polyclinic's address \\
\texttt{Name}                & Official name of the polyclinic \\
\texttt{ZIP Code}            & Postal (ZIP) code of the polyclinic \\
\texttt{Country}             & Country where the polyclinic is located \\
\texttt{State}               & State where the polyclinic is located \\
\texttt{Health Region}       & Administrative health region the polyclinic serves \\
\texttt{CNES Code}           & National Registry of Health Establishments (CNES) code \\
\texttt{Address Complement}  & Additional address details (e.g., floor, suite) \\
\texttt{Generic Name}        & Common or abbreviated name of the polyclinic \\
\bottomrule
\end{tabular}
\caption{Polyclinics Dataset Schema.}
\label{tab:polyclinics_schema}
\end{table}

\noindent \textbf{Hospitals}: compiles the public and private hospital facilities operating within the municipality, providing inpatient care, emergency services, and complex medical procedures; hospitals represent the tertiary tier of Niterói's health network and serve residents from across the municipality and neighboring areas. Table~\ref{tab:hospitals_schema} presents the key variables of the hospitals dataset and their descriptions.

\begin{table}[htb]
\centering
\begin{tabular}{lp{10cm}}
\toprule
\textbf{Variable} & \textbf{Description} \\
\midrule
\texttt{X Coordinate}      & Projected X coordinate (easting) of the hospital's location \\
\texttt{Y Coordinate}      & Projected Y coordinate (northing) of the hospital's location \\
\texttt{Object ID}         & Unique numeric identifier for each record \\
\texttt{Name Text}         & Official name of the hospital \\
\texttt{Address Text}      & Street name and number of the hospital's address \\
\texttt{ZIP Code Text}     & Postal (ZIP) code of the hospital \\
\texttt{Neighborhood Text} & Neighborhood where the hospital is located \\
\texttt{Website Text}      & Official website URL \\
\texttt{Email Text}        & Official contact email address \\
\texttt{Telephone Text}    & Contact telephone number \\
\bottomrule
\end{tabular}
\caption{Hospitals Dataset Schema.}
\label{tab:hospitals_schema}
\end{table}

\noindent \textbf{Pharmacies}: inventories the licensed pharmacies and drugstores operating throughout the municipality's neighborhoods, encompassing both independent establishments and chain outlets; pharmacies play a key role in the health system by dispensing medications, including those distributed free of charge through public health programs. Table~\ref{tab:pharmacies_schema} presents the key variables of the pharmacies dataset and their descriptions.

\begin{table}[htb]
\centering
\begin{tabular}{lp{10cm}}
\toprule
\textbf{Variable} & \textbf{Description} \\
\midrule
\texttt{X Coordinate}        & Projected X coordinate (easting) of the pharmacy's location \\
\texttt{Y Coordinate}        & Projected Y coordinate (northing) of the pharmacy's location \\
\texttt{Object ID}           & Unique numeric identifier for each record \\
\texttt{Name Text}           & Official or commercial name of the pharmacy \\
\texttt{Address Text}        & Street name of the pharmacy's address \\
\texttt{Address Number Text} & Street number of the pharmacy's address \\
\texttt{Neighborhood Text}   & Neighborhood where the pharmacy is located \\
\bottomrule
\end{tabular}
\caption{Pharmacies Dataset Schema.}
\label{tab:pharmacies_schema}
\end{table}

The \NetMob \textbf{Mobility} dataset of Niterói comprises three mobility datasets, each capturing a distinct layer of the active and vehicular transportation infrastructure across the neighborhoods.

\noindent \textbf{Bicycle Stations}: catalogs the public bicycle-sharing stations deployed across Niterói as part of the municipality's active mobility program, recording each station's physical location, structural model, capacity, and implementation date; the network supports short-distance trips and last-mile connectivity, complementing public transit routes along the waterfront and main corridors. Table~\ref{tab:bicycle_stations_schema} presents the key variables of the bicycle stations dataset and their descriptions.

\begin{table}[htb]
\centering
\begin{tabular}{lp{10cm}}
\toprule
\textbf{Variable} & \textbf{Description} \\
\midrule
\texttt{X Coordinate}       & Longitude coordinate of the station's location \\
\texttt{Y Coordinate}       & Latitude coordinate of the station's location \\
\texttt{Object ID}          & Unique numeric identifier for each record \\
\texttt{Number List}        & Number of bicycle docks available at the station \\
\texttt{Model Text}         & Bicycle model type available at the station (e.g., Adulto) \\
\texttt{Name Text}          & Official name of the bicycle station \\
\texttt{Location Text}      & Descriptive location or reference point of the station \\
\texttt{Type Text}          & Physical layout type of the station (e.g., Dupla, Linear) \\
\texttt{Implementation Date}& Date on which the station was installed and made operational \\
\bottomrule
\end{tabular}
\caption{Bicycle Stations Dataset Schema.}
\label{tab:bicycle_stations_schema}
\end{table}

\noindent \textbf{Private Parking Lots}: inventories the privately operated off-street parking facilities available to the public across Niterói's neighborhoods, including standalone garages and lots managed by national and local operators; these facilities complement on-street parking supply and are concentrated in commercial and high-density residential areas. Table~\ref{tab:private_parking_schema} presents the key variables of the private parking lots dataset and their descriptions.

\begin{table}[htb]
\centering
\begin{tabular}{lp{10cm}}
\toprule
\textbf{Variable} & \textbf{Description} \\
\midrule
\texttt{X Coordinate}      & Projected X coordinate (easting) of the parking lot's location \\
\texttt{Y Coordinate}      & Projected Y coordinate (northing) of the parking lot's location \\
\texttt{Object ID}         & Primary unique numeric identifier for each record \\
\texttt{Object ID 1}       & Secondary numeric identifier for each record \\
\texttt{Name Text}         & Official or commercial name of the parking facility \\
\texttt{Telephone Text}    & Contact telephone number \\
\texttt{Email Text}        & Official contact email address \\
\texttt{Address Text}      & Full street address of the parking facility \\
\texttt{ZIP Code Text}     & Postal (ZIP) code of the parking facility \\
\texttt{Website Text}      & Official website URL \\
\texttt{Neighborhood Text} & Neighborhood where the parking facility is located \\
\bottomrule
\end{tabular}
\caption{Private Parking Lots Dataset Schema.}
\label{tab:private_parking_schema}
\end{table}

\noindent \textbf{Rotating Parking Spaces} (Zona Azul): maps the individual on-street regulated parking spaces that form part of Niterói's paid rotating parking scheme, recording the precise location, street geometry, vacancy type, and operational rules for each space; the rotating system is designed to increase parking turnover in high-demand commercial areas, and is organized by sector and administrative region. Table~\ref{tab:rotating_parking_schema} presents the key variables of the rotating parking spaces dataset and their descriptions.

\begin{table}[h!]
\centering
\begin{tabular}{lp{10cm}}
\toprule
\textbf{Variable} & \textbf{Description} \\
\midrule
\texttt{X Coordinate}             & Projected X coordinate (easting) of the parking space's location \\
\texttt{Y Coordinate}             & Projected Y coordinate (northing) of the parking space's location \\
\texttt{Object ID}                & Unique numeric identifier for each record \\
\texttt{Global Vacancy Number Text} & Global sequential vacancy number within the regulated zone \\
\texttt{Initial Coordinate Text}  & Starting coordinate reference along the street segment \\
\texttt{Final Coordinate Text}    & Ending coordinate reference along the street segment \\
\texttt{Database Latitude}        & Latitude coordinate stored in the source database \\
\texttt{Database Longitude}       & Longitude coordinate stored in the source database \\
\texttt{Street Type Text}         & Traffic flow type of the street (e.g., Mão Única) \\
\texttt{Face Text}                & Side of the street where the space is located  \\
\texttt{Vacancy Type Text}        & Category of the parking space (e.g., Convencional, Especial) \\
\texttt{Rule Text}                & Operational rule or restriction applied to the space \\
\texttt{Area Text}                & Named regulated parking area the space belongs to \\
\texttt{Region Text}              & Administrative region associated with the space \\
\texttt{Sector Text}              & Numeric sector code within the regulated parking scheme \\
\texttt{Street Name Text}         & Name of the street where the space is located \\
\texttt{Numeral Type Text}        & Type of street numbering reference used (e.g., Número) \\
\texttt{Street Number Text}       & Street number nearest to the parking space \\
\texttt{Neighborhoods Text}       & Neighborhood where the space is located \\
\texttt{Administrative Region Text} & Broader administrative region of the municipality \\
\texttt{Reference Text}           & Additional spatial reference or landmark description \\
\texttt{Vacancy Text}             & Orientation of the space relative to the kerb  \\
\texttt{Corner Text}              & Position of the space relative to a street corner, if applicable \\
\texttt{Global ID}                & Globally unique identifier (GUID) for the record \\
\bottomrule
\end{tabular}
\caption{Rotating Parking Spaces Dataset Schema.}
\label{tab:rotating_parking_schema}
\end{table}

The \NetMob \textbf{Garbage Collection Schedule} dataset of Niterói documents the municipal solid waste collection service across the city's neighborhoods, mapping each collection zone to its route code, operating schedule, and spatial geometry. The dataset covers 63 collection zones distributed across five route series — RPB (Praias da Baía), RN (Norte), RO (Oceânica), RP (Pendotiba), and RL (Leste) — reflecting the administrative regions into which waste management is organized. Three distinct daytime schedules are in operation: Monday/Wednesday/Friday from 7:00 to 15:00 (18 zones), Tuesday/Thursday/Saturday from 7:00 to 15:00 (15 zones), and a daily nocturnal service starting at 20:00 (22 zones), primarily covering the denser Praias da Baía coastal area; eight zones lack schedule information. Some neighborhoods are split across multiple collection polygons, including island territories such as Ilha do Pai, Ilha da Mãe, and Ilha da Conceição, each recorded as separate spatial units. Table~\ref{tab:garbage_collection_schema} presents the key variables of the garbage collection schedule dataset and their descriptions.

\begin{table}[h!]
\centering
\begin{tabular}{lp{10cm}}
\toprule
\textbf{Variable} & \textbf{Description} \\
\midrule
\texttt{Object ID}                  & Unique numeric identifier for each collection zone record \\
\texttt{Neighborhood Text}          & Name of the neighborhood corresponding to the collection zone \\
\texttt{Code Text}                  & Route code of the collection zone (e.g., RPB05, RN01, RO09), indicating the administrative route series and sequence number \\
\texttt{Notes Text}                 & Additional spatial notes for the zone, such as island or locality name when a neighborhood spans multiple polygons \\
\texttt{Database Area}              & Area of the collection zone polygon as stored in the source database (in square kilometres) \\
\texttt{Collection Datetime Text}   & Operating schedule and time window for waste collection in the zone (e.g., Segunda/Quarta/Sexta de 7:00 às 15:00; Diária noturna a partir de 20:00) \\
\texttt{Global ID}                  & Globally unique identifier (GUID) for the record \\
\texttt{Shape Length}               & Perimeter length of the collection zone polygon (in projected units) \\
\texttt{Shape Area}                 & Surface area of the collection zone polygon (in projected units) \\
\bottomrule
\end{tabular}
\caption{Garbage Collection Schedule Dataset Schema.}
\label{tab:garbage_collection_schema}
\end{table}

The \NetMob \textbf{Restaurants} dataset of Niterói inventories the food and beverage establishments operating across the municipality as of 2019, providing geocoded records for 1{,}268 venues distributed across 51 neighborhoods. The dataset spans 23 establishment categories. The geographic concentration is highest in Centro (237 venues) and Icaraí (207), which together account for more than a third of all recorded establishments. Table~\ref{tab:restaurants_schema} presents the key variables of the restaurants dataset and their descriptions.

\begin{table}[h!]
\centering
\begin{tabular}{lp{10cm}}
\toprule
\textbf{Variable} & \textbf{Description} \\
\midrule
\texttt{X Coordinate}        & Projected X coordinate (easting) of the establishment's location \\
\texttt{Y Coordinate}        & Projected Y coordinate (northing) of the establishment's location \\
\texttt{Object ID}           & Unique numeric identifier for each record \\
\texttt{Name Text}           & Official or commercial name of the food and beverage establishment \\
\texttt{Category Text}       & Type of food service establishment (e.g., Cardápio/Buffet, Bar, Lanchonete, Pizza, Padaria) \\
\texttt{Address Text}        & Street name of the establishment's address \\
\texttt{Address Number Text} & Street number of the establishment's address \\
\texttt{Neighborhood Text}   & Neighborhood where the establishment is located \\
\texttt{ZIP Code Text}       & Postal (ZIP) code of the establishment \\
\texttt{Telephone Text}      & Contact telephone number \\
\texttt{Email Text}          & Official contact email address \\
\texttt{Website Text}        & Official website URL \\
\bottomrule
\end{tabular}
\caption{Restaurants Dataset Schema.}
\label{tab:restaurants_schema}
\end{table}

\end{document}